\documentclass[symmetry,article,submit,moreauthors,pdftex,12pt,a4paper]{mdpi} 
\lastpage{x}
\doinum{10.3390/------}
\pubvolume{xx}
\pubyear{2010}
\history{Received: xx / Accepted: xx / Published: xx}
\pdfoutput=1

\newcommand{\be}{\begin{equation}}
\newcommand{\ee}{\end{equation}}
\newcommand{\bea}{\begin{eqnarray}}
\newcommand{\eea}{\end{eqnarray}}
\newcommand{\ba}{\begin{array}}
\newcommand{\ea}{\end{array}}
\newcommand{\lr}{\leftrightarrow}

\usepackage{amssymb,amsmath}
\usepackage{graphicx}
\usepackage{epsfig}
\usepackage{caption}
\Title{Complex Networks and Symmetry II: Reciprocity and Evolution
of World Trade}

\Author{Franco Ruzzenenti $^{1,\star}$, Diego Garlaschelli $^{2}$ and Riccardo Basosi $^{3}$}

\address{%
$^{1}$ Center for the Study of Complex Systems, University of Siena, Via Roma 56, 53100 Siena, Italy\\
$^{2}$ LEM, Sant'Anna School of Advanced Studies, P.zza Martiri della Libert\`a 33, 56127 Pisa, Italy; \linebreak E-Mail: garlaschelli.diego@gmail.com\\
$^{3}$ Department of Chemistry, University of Siena, Via Aldo Moro
1, 53100 Siena, Italy; \linebreak E-Mail:basosi@unisi.it}

\corres{E-Mail: ruzzenenti@unisi.it; \linebreak  Tel.: +39-0577-234240; Fax:
+39-0577-234239.}

\abstract{We exploit the symmetry concepts developed in the
companion review of this article to introduce a stochastic version
of link reversal symmetry, which leads to an improved understanding
of the reciprocity of directed networks. We apply our formalism to
the international trade network and show that a strong embedding in
economic space determines particular symmetries of the network,
while the observed evolution of reciprocity is consistent with a
symmetry breaking taking place in production space. Our results show
that networks can be strongly affected by symmetry-breaking
phenomena occurring in embedding spaces, and that stochastic network
symmetries can successfully suggest, or rule out, possible
underlying mechanisms. }


\begin{document}
\section{Introduction}
In this paper, we take full advantage to the symmetry concepts
developed in the companion \linebreak review \cite{symmetry1} in
order to study in great detail two applications of stochastic
symmetry in networks. First, we discuss link reversal symmetry in
directed networks and introduce its stochastic variant to highlight
the connections with the important property of reciprocity. Second,
we consider the problem of (stochastic) symmetry breaking in
spatially embedded networks (where the embedding space is in general
not necessarily Euclidean or geographic, but specified by different
variables). Both applications will be discussed in the context of an
empirical analysis of a particular network: the World Trade Web,
defined by the international trade relationships existing among all
world countries.

The investigation of symmetry naturally leads to the fascinating
problem of symmetry breaking. `All scientific applications of
symmetry are based on the principle that \emph{identical causes
produce identical effects}' \cite{1}. That is to say, the symmetry
of the effect must be at least that of the cause, or in a
mathematical jargon: the order of the \emph{symmetry group} of the
effect must be at least equivalent to that of the cause.
Nevertheless, for a qualitatively new phenomenon to occur, symmetry
cannot be conserved. Pierre Curie was the first in modern science
who highlighted the relevance of spontaneous symmetry breaking in
various phenomena \cite{2}. His studies of criticality in phase
transitions overcame the boundaries of solid state physics and posed
a suitable analytical framework for further studies, in different
fields. Prigogine himself, a renowned and illustrious precursor of
complex systems research, referred to Curie's contribution in order
to elucidate the meaning of symmetry breaking in dissipative
structures: `We see therefore, that the appearance of a periodic
reaction is a time-symmetry breaking process exactly as
ferromagnetism is a space-symmetry breaking one' \cite{3}. Two main
viewpoints on symmetry breaking developed in science: one concerned
with space symmetry and one with time symmetry. Symmetry breaking
can be indeed approached from two different sides: from a time-scale
perspective, as we see the phenomenon as a dynamic system, or from a
spatial perspective, as we focus on changes in the system's space.
With the former approach we tend to consider changes that are
endogenous to the system, while space is taken as homogeneous all
the way through; with the latter, the system is embedded in some
space and changes are considered exogenous. However, whereas in
physics space-symmetry breaking has been a prominent research
subject, in the field of complex nonlinear systems time-symmetry
breaking has received much more attention. The following passage by
Mainzer illustrates on what basis time symmetry breaking became a
major topic in the science of complexity: `Thus, bifurcation
mathematically only means the emergence of new solutions of
equations at critical values. Actually, bifurcation and symmetry
breaking is a purely mathematical consequence of the theory of
nonlinear differential equations. But, bifurcations of final states
as solutions of differential equations correspond to qualitative
changes of dynamical systems and the emergence of new phenomena in
nature and society [..]'\cite{4}.

In what follows, spatial symmetry breaking will be approached in the
framework of complex networks. Symmetries relevant to networks can
be either `internal', if they involve purely topological quantities,
or `external', if they are defined with respect to additional
properties such as positions in some embedding space. In the latter
case,  symmetries relative to the external space can be reflected in
some topological property displayed by the network \cite{symmetry1}.
In this sense, spatial symmetry breaking has so far received little
attention in the field of network theory, despite the latter
developed considerably in recent years
\cite{guidosbook,largescalestructure,
dynamicalprocessesoncomplexnetworks,
ecologicalnetworks,internet,networksincellbiology,adaptivenetworks}.
On the other hand, specific analyses of processes that are well
described within a network framework suggest that spatial symmetry
breaking can occur with respect to some embedding space and manifest
itself in major structural changes at a topological level, as
happened in the evolution of vascular systems in living beings
\cite{west}, of river basins \cite{banavar}, and of production
networks in modern economies \cite{7,8}. In the present paper, we
exploit our review of network symmetries \cite{symmetry1} to start
investigating this problem. The paper is organized as follows. In
Section \ref{sec_exact} we will introduce link reversal symmetry and
we define a new stochastic variant of it. In Section
\ref{sec_reciprocity} we will then  investigate how stochastic link
reversal symmetry, together with other symmetries discussed in
referenc\cite{symmetry1}, enable to achieve an improved
understanding of network structure in a specific case, {\em i.e.} the
problem of \emph{reciprocity} in directed networks. We will
highlight how different measures of reciprocity capture different
symmetry properties of a network. This will help us disentangle
distinct possible mechanisms explaining the observed reciprocity
structure of real networks. As a particular application, in Section
\ref{sec_wtw} we will consider the evolution of reciprocity in the
World Trade Web. We will also emphasize the role of spatial
embedding, which relates the topology of the network to underlying
geographical coordinates and economic variables. We will advance
heuristic explanations for the observed evolution of reciprocity in
the World Trade Web in terms of symmetry breaking phenomena due to
changes in the underlying economic structure. These analyses
highlights the idea that complex networks are not phenomena
\emph{per se}, but maps of physical phenomena that are immersed in
physical space---or any other space, depending on the variables
determining the system's dynamics. Symmetry breaking can occur in
some geographical, economic, or different space, and be mirrored in
the topological space the network belongs to. In real imperfect
systems, stochastic symmetry is able to capture spatial patterns
that are undetected by exact symmetries. Interactions between the
underlying system's `spaces' is an intriguing challenge for network
theory and pertains the study of network dynamics.

\section{Exact and Stochastic Link Reversal Symmetry\label{sec_exact}}
In this section we make use of the notion of \emph{stochastic graph
symmetries} we introduced in \linebreak Reference \cite{symmetry1}
to define a new graph invariance, {\em i.e.} the stochastic version
of \emph{link reversal symmetry}. To this end, we first briefly
recall the concept of graph ensembles and equiprobability
\cite{symmetry1}, and then discuss link reversal symmetry, first in
its exact version and finally in its stochastic variant.

\subsection{Graph Ensembles and Stochastic Symmetries\label{sec_ensembles}}
In Reference \cite{symmetry1} we introduced the concept of graph
ensembles as collections of graphs with specified properties and
probability. Each graph $G$ in a statistical ensemble of graphs has
an associated occurrence probability $P(G)$ satisfying
\begin{equation}
\sum_G P(G)=1
\end{equation}
Two graphs $G_1$ and $G_2$ such that $P(G_1)=P(G_2)$ are said to be
\emph{equiprobable} in the ensemble considered. The probability
$P(G)$ can have different forms depending on the structure of the
ensemble under consideration. In what follows, we will make use of
\emph{(grand)canonical} ensembles, and in particular maximally
random graphs with specified constraints \cite{symmetry1}. Such
ensembles are defined by specifying the expected value (ensemble
average) of a chosen set of topological properties (the
constraints), and are maximally random otherwise. This means that
the probability $P(G)$ must maximize the entropy of the ensemble
subject to the enforced constraints. The constraints will be denoted
as a collection $\{c_1,\dots, c_K\}$ of $K$ topological properties,
and the Lagrange multipliers involved in the constrained
maximization problem will be denoted as the conjugate parameters
$\{\theta_1,\dots, \theta_K\}$. The probability $P(G)$ will depend
on such parameters, and its explicit form is \be
P(G)=\frac{e^{-H(G)}}{Z} \ee where $H(G)$ is the \emph{graph
Hamiltonian} \be H(G)\equiv \sum_{a=1}^K \theta_a c_a(G) \ee and $Z$
is the \emph{partition function} \be Z\equiv\sum_G e^{-H(G)} \ee The
expected value of a topological property $X$, which evaluates to
$X(G)$ on the particular graph $G$, is
\begin{equation}
\langle X(\theta_1,\dots, \theta_K)\rangle\equiv\sum_G P(G) X(G)
\label{eq_X}
\end{equation}
The values of the parameters $\{\theta_1,\dots, \theta_K\}$ are such
that the expected values $\{\langle c_1\rangle,\dots, \langle
c_K\rangle\}$ of the constraints match the specified values. In
particular, if the ensemble is meant as a null model
\cite{symmetry1} of a real network $G^*$, the expected values of the
constraints will have to match the empirical values of the
properties $\{c_1(G^*),\dots, c_K(G^*)\}$ of that particular graph:
\begin{equation}
\langle c_a(\theta^*_1,\dots, \theta^*_K)\rangle=c_a(G^*)\qquad a=1,\dots,K
\end{equation}
The above parameter choice automatically maximises the probability
$P(G^*)$ to obtain the real network $G^*$ under the model
considered, and is therefore in accordance with the \emph{maximum
likelihood \linebreak principle} \cite{mylikelihood}. The graph
Hamiltonian $H(G)$, which represents a sort of energy or cost
associated to the graph $G$, is a linear combination of the
constraints. Clearly, $P(G_1)=P(G_2)$ if and only if
$H(G_1)=H(G_2)$, which means that graphs with the same energy are
equiprobable (and vice versa). The transformation mapping $G_1$ into
a different graph $G_2$ with $H(G_2)=H(G_1)$ is a symmetry of the
Hamiltonian. Any such transformation changes the topology of the
graph but preserves the values of the constraints appearing in the
Hamiltonian (and is therefore more general than permutations of
vertices).

Graph ensembles provide an ideal framework to study stochastic graph
symmetries, that we defined in Reference \cite{symmetry1}. An exact
symmetry of a real network $G^*$ is a transformation mapping $G^*$
to itself (for instance, an automorphism if the transformation
considered is a vertex permutation
\cite{symmetry,quotient,redundancy,symmetry_wtw}). By contrast, a
stochastic symmetry is associated with an ensemble of graphs, rather
than with a single one. In particular, we can define a graph
ensemble as stochastically symmetric under a transformation if the
latter maps each graph $G_1$ into an equiprobable subgraph $G_2$
with $P(G_1)=P(G_2)$. Maximally random graphs with constraints are
therefore stochastically symmetric under  transformations that are
symmetries of the Hamiltonian. If a real network $G^*$ is well
reproduced by a stochastically symmetric ensemble, then we can
denote $G^*$ as stochastically symmetric (under the same
transformations involved in the symmetry of the ensemble, or
\emph{under the model considered} for brevity). That is, while $G^*$
is exactly symmetric under its automorphisms, it is stochastically
symmetric under the transformations defining an ensemble of which
$G^*$ is a typical member. By contrast, if the ensemble is not a
good model of the real network,then those transformations are not
stochastic symmetries of $G^*$. We will encounter both situations
later on.

Two important examples of maximally random graphs that we will use
as null models are the Erd\H{o}s-R\'enyi random graph model and the
configuration model. Here we consider the undirected versions of
both models, and we will generalize them to the directed case later
on. To avoid confusion with their directed counterparts, here we use
a different notation with respect to our presentation in Reference
\cite{symmetry1}. The adjacency matrix of an undirected graph will
be denoted as $B$, with entries $b_{ij}=1$ if an undirected link
between vertices $i$ and $j$ is there, and $b_{ij}=0$ otherwise.

In the undirected Erd\H{o}s-R\'enyi random graph model, the only
constraint is the total number of undirected links
$L^u=\sum_{i<j}b_{ij}$. Thus the Hamiltonian reads
\begin{equation}
H(G)=\theta L^u(G)
\label{eq_Hrand}
\end{equation}
and its symmetries are the transformations mapping a graph $G$ into
another (equiprobable) graph with the same number of links. The
probability $P(G)$ factorizes in terms of the probability
\begin{equation}
q\equiv \frac{e^{-\theta}}{1+e^{-\theta}}
\end{equation}
that a link is there between any two vertices. If the model is
interpreted as a null model of the real network $G^*$, the parameter
$q$ must be set to the particular value $q^*$ such that
\begin{equation}
\langle L^u\rangle=q^*\frac{N(N-1)}{2}=L^u(G^*)
\label{eq_likelihoodL}
\end{equation}
ensuring that, in accordance with the maximum likelihood principle
\cite{mylikelihood}, the expected number of links $\langle
L^u\rangle$ coincides with the number of links $L^u(G^*)$ of $G^*$.

In the configuration model, the constraints are the degrees of all
vertices, {\em i.e.} the \emph{degree sequence} $\{k_i\}$, where
$k_i\equiv\sum_{j\ne i}b_{ij}$. Therefore the Hamiltonian takes the
form \be H(G)=\sum_{i=1}^N \theta_i k_i(G) \label{eq_Hconf} \ee and
its symmetries are the transformations that map a graph into a
different one with the same degree sequence, {\em i.e.} those
explored by the \emph{local rewiring algorithm} \cite{symmetry1}.
The probability that vertices $i$ and $j$ are connected is no longer
uniform across all pairs of vertices, and reads
\begin{equation}
q_{ij}=\frac{w_i w_j}{1+w_i w_j}
\label{eq_pij}
\end{equation}
where $w_i\equiv e^{-\theta_i}$. If the configuration model is used
as a null model of a real network $G^*$, then the parameters
$\{w_1,\dots,w_N\}$ must be set to the values
$\{w^*_1,\dots,w^*_N\}$  solving the following $N$ coupled equations
\begin{equation}
\langle k_i\rangle=\sum_{j\ne i}\frac{w^*_i w^*_j}{1+w^*_i w^*_j}=k_i(G^*)\qquad \forall i
\label{eq_likelihoodk}
\end{equation}
ensuring that the expected degree sequence coincides with the
observed one, and maximising the likelihood to obtain $G^*$
\cite{mylikelihood,myrandomization}.

\subsection{Transpose Equivalence and Transpose Equiprobability\label{sec_reversal}}
We now come to the description of \emph{link reversal} symmetry.
There are two ways in which one can formulate link reversal symmetry
in directed networks. The first, simpler definition is the exact
invariance of a single graph under the inversion of the direction
defined on each of its edges. Under this definition, the graph is
perfectly symmetric if all of its edges are bidirectional. If $A$ is
the adjacency matrix of a directed graph ($a_{ij}=1$ if a directed
edge from $i$ to $j$ is there, and $a_{ij}=0$ otherwise), then the
graph is exactly symmetric under link reversal if
\begin{equation}
A=A^T
\end{equation}
where $A^T$ indicates the transpose of the matrix $A$. Clearly,
bidirectional graphs are equivalent to undirected graphs. In this
sense, one can say that real networks are found to be either
symmetric (this is the case of real-world undirected networks such
as the Internet, protein interaction graphs or friendship networks)
or asymmetric (this is the case of intrinsically directed networks
such as food webs, the WWW, metabolic networks, the World Trade Web,
{\em etc.}). This first type of link reversal symmetry will be
denoted \emph{transpose equivalence} in what follows.

A second, novel definition of link reversal symmetry that we
introduce here is a stochastic one, in the sense discussed in
Reference \cite{symmetry1} and briefly recalled above. As any
stochastic symmetry, it is associated to an ensemble of equiprobable
graphs. If each graph $G$ in the ensemble is identified with its
adjacency matrix $A$, we say that the ensemble is stochastically
symmetric under link reversal if
\begin{equation}
P(A)=P(A^T)
\label{eq_PP}
\end{equation}
This second definition is completely different from the first one.
It does not imply that any single graph $A$ in the ensemble is
bidirectional, but that it has the same probability of occurrence of
its link-reversed  $A^T$, {\em i.e.} $P(A)=P(A^T)$. The
equiprobability of $A$ and $A^T$ has important effects on the
directionality of the expected topological properties across the
ensemble, but is perfectly consistent with the asymmetry of
individual graphs in the ensemble. If the ensemble considered is a
maximally random graph model defined by a Hamiltonian $H(G)$ (see
Section \ref{sec_ensembles}), then Equation (\ref{eq_PP}) is
equivalent to
\begin{equation}
H(A)=H(A^T)
\end{equation}
showing that link reversal is a symmetry of the Hamiltonian. In
accordance with our general definition of stochastic symmetry
\cite{symmetry1} recalled in Section \ref{sec_ensembles}, we can
also define a single graph $G^*$ as stochastically symmetric under
link reversal if it is a typical member of ({\em i.e.} it is well
modelled by) an ensemble which is  stochastically symmetric under
link reversal. In simpler words, the graph $A$ is stochastically
symmetric under link reversal if it is statistically equivalent to
its link-reversed $A^T$. This second, stochastic type of link
reversal symmetry will be denoted \emph{transpose equiprobability}
in what follows.

The dichotomy existing between transpose equivalence and transpose
equiprobability, the different underlying mechanisms they might
reveal, and the relation they have to many of the symmetries we have
discussed in Reference \cite{symmetry1} (including ensemble
equiprobability, statistical equivalence and dependence on external
or hidden vertex properties) make link reversal symmetry an ideal
candidate to discuss in more detail in what follows. Moreover, link
reversal symmetry is tightly related to the problem of
\emph{reciprocity}. Therefore, before presenting a deeper study of
this symmetry, in the next section we study the problem of
reciprocity in great detail.

\section{Reciprocity of Directed Networks\label{sec_reciprocity}}
Reciprocity is the tendency of pairs of vertices to be connected by
two mutual links pointing in opposite directions, a particular type
of correlation found in directed networks
\cite{wasserman,myreciprocity,mymultispecies}. Depending on the
nature of the network, reciprocity is related to various important
phenomena, such as ecological symbiosis in food webs, reversibility
of biochemical reactions in metabolic networks, bidirectionality of
chemical synapses in neural networks, synonymy in networks of
dictionary terms, mutuality of psychological associations in
networks of freely linked words, reciprocity of hyperlinks in the
WWW, crossed financial ownership in shareholding networks, economic
interdependence of countries in the international trade network, and
so on \cite{myreciprocity}. In this section, we study link reversal
symmetry in great detail. We first discuss the  problem of the
definition of proper reciprocity measures, present the analysis of
the reciprocity structure of real networks, and define some
theoretical concepts useful to interpret the observed patterns.
Then, in Section \ref{sec_relation} we highlight the relation
existing between reciprocity, the two types of link reversal
symmetry defined in Section \ref{sec_reversal}, and other symmetries
we introduced. \linebreak In Section \ref{sec_wtw} we finally apply
all these concepts to the empirical analysis of the World Trade Web.

\subsection{The Traditional Approach to Reciprocity}\label{sec_link}
\begin{figure}[h]
\begin{center}
\includegraphics[width=0.6\textwidth]{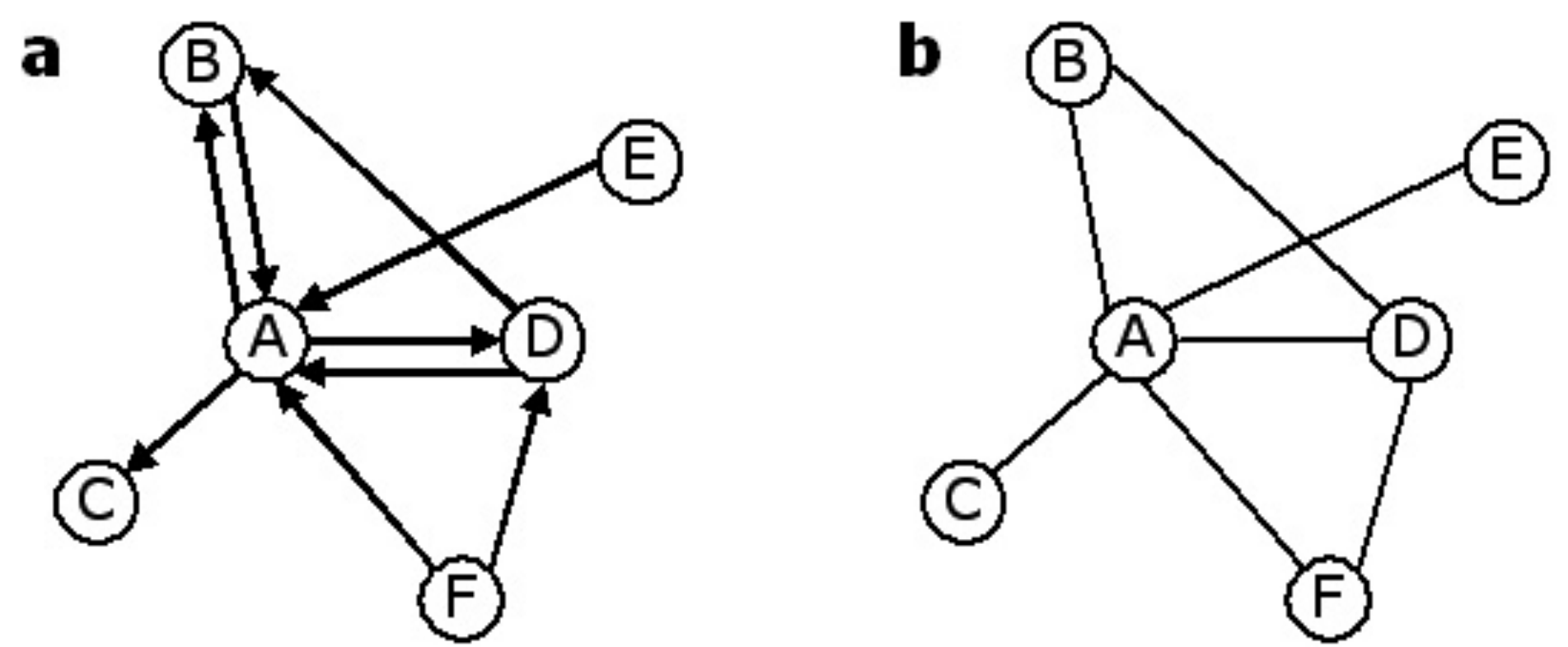}
\end{center}
\caption{(a) Example of a directed network with $N=6$ vertices. Here
$L=9$, $L^\lr=4$ and the maximum possible number of directed links
is $N(N-1)=30$. (b) The undirected version of the same network. Here
$L^u=7$ and the maximum possible number of undirected links is
$N(N-1)/2=15$. \label{fig_dirundgraph}}
\end{figure}

The study of reciprocity has a long tradition in social science
\cite{wasserman} as a way to quantify how many `ties' (directed
links) are reciprocated in a social network of `actors' (vertices).
The \emph{reciprocal link} of a directed link pointing from $i$ to
$j$ is a link pointing from $j$ to $i$. A link is
\emph{reciprocated} if its reciprocal one is present in the network.
In terms of the adjacency matrix of the graph, two reciprocated
links are present between $i$ and $j$ if and only if
$a_{ij}=a_{ji}=1$. In the example shown in Figure
\ref{fig_dirundgraph}a, the edges between vertices $A$ and $B$, as
well as those between $A$ and $D$, are reciprocated. All other edges
are not reciprocated. Therefore, while the total number of directed
links is given by \be\label{L} L=\sum_{i\ne j}a_{ij} \ee the number
of reciprocated links is \be\label{Lboth} L^\lr=\sum_{i\ne
j}a_{ij}a_{ji} \ee Since $0\le L^\lr\le L$, the traditional
definition of the reciprocity of a network is \be
r\equiv\frac{L^\lr}{L} \label{r} \ee so that $0\le r\le 1$. Although
not usually remarked, it is important to notice that whether the
value of $r$ can actually span the entire range between $0$ and $1$
depends on the link density (or \emph{connectance}) of the network,
defined as \be \bar{a}\equiv\frac{\sum_{i\ne
j}a_{ij}}{N(N-1)}=\frac{L}{N(N-1)} \label{eq_conn} \ee We shall
comment more about the effects of $\bar{a}$ on the allowed values of
the reciprocity later on. Note that the requirement $i\ne j$ in
Equations (\ref{L}), (\ref{Lboth}) and (\ref{eq_conn}) arises from
the assumption of no self-loops (links starting and ending at the
same vertex) in the network. If self-loops are present, we assume
that they are ignored and therefore not computed in $L$ and $L^\lr$.
This is because self-loops would give a nonzero contribution to both
$L$ and $L^\lr$, even if they are not a true signature of
reciprocity. Two networks with the same topology apart from a
different number of self-loops should not be considered as having
different degrees of reciprocity \cite{myreciprocity}.

As for any topological property, a given value of $r$ is only
significant with respect to some null model. This is because, even
in a network where directed links are drawn completely at random, a
certain number of reciprocated connections will be formed. As we
shall discuss in more detail in Section \ref{sec_relation}, in such
an uncorrelated network $r$ is simply equal to the average
probability that \emph{any} two vertices are connected by a directed
link, {\em i.e.} to the connectance defined in Equation
(\ref{eq_conn}): \be r_{rand}=\bar{a} \label{rand} \ee Comparing the
value of $r$ with that of $r_{rand}$ allows to assess if mutual
links occur more ($r>r_{rand}$) or less ($r<r_{rand}$) often than
expected by chance. This is the traditional approach to the study of
reciprocity in social networks, which has been more recently
extended to other networks such as the WWW, e-mail networks and the
World Trade Web \cite{myreciprocity}.

\subsection{An Improved Definition\label{sec_rho}}
Although the comparison of $r$ with $r_{rand}$ is a safe method to
detect nonrandom reciprocity in a particular network, it is
completely unadapted to compare the  reciprocity of networks with
different link density, or to assess the evolution of reciprocity in
a single network with time-varying density \cite{myreciprocity}.
This is because $r$ is not an absolute quantity, and its value has
only a relative meaning with respect to $r_{rand}=\bar{a}$. The
reference value for $r$ unavoidably varies as the density $\bar{a}$
varies. Therefore it is not possible to order various networks, or
various snapshots of the same network, according to their value
\linebreak of $r$. In order to overcome this problem, a new
definition of reciprocity was proposed \cite{myreciprocity} as the
Pearson correlation coefficient between the symmetric entries of the
adjacency matrix: \be\label{rho} \rho\equiv\frac{\sum_{i\ne j}
(a_{ij}-\bar{a})(a_{ji}-\bar{a})} {\sum_{i\ne j}(a_{ij}-\bar{a})^2}
=\frac{r-\bar{a}}{1-\bar{a}} \ee where the second equality comes
from an explicit calculation making use of Equations (\ref{L})--
(\ref{r}) and of the property ${(a_{ij})^2=a_{ij}}$. The range of
$\rho$, as for any correlation coefficient, is $-1\le \rho\le 1$
(see however our discussion below for more details on the allowed
values of $\rho$). It is possible to write down an expression for
the statistical error associated to a single measurement of $\rho$
on a particular \linebreak network \cite{myreciprocity}.

Unlike $r$, $\rho$ is an absolute quantity, and the effects of link
density are already accounted for in it. In particular, its null
value is \be \rho_{rand}=0 \label{rhorand} \ee irrespective of the
value of $\bar{a}$. The sign of $\rho$ alone is enough to
distinguish between positively correlated (or \emph{reciprocal})
networks where there are more reciprocated links than expected by
chance ($\rho>0$) and  negatively correlated (or
\emph{antireciprocal}) networks where there are fewer reciprocated
links than expected by chance ($\rho<0$). The null case $\rho=0$
(consistently with the statistical error) corresponds to
uncorrelated or \emph{areciprocal} networks. The existence of a
unique reference scale  allows to order several networks according
to their value of $\rho$, as shown in Table \ref{tab}. Among the
networks considered, one finds both positively and and negatively
correlated ones. Remarkably, such ordering reveals interesting
empirical patterns of reciprocity, since networks of the same kind
are found to display similar values \linebreak of $\rho$. The positively
correlated networks are, in decreasing order of $\rho$ (see Table
\ref{tab}): all purely bidirectional (undirected) networks such as
the Internet ($\rho=1$), the 53 snapshots of the World Trade Web
from year 1948 to 2000 ($0.68\le\rho\le 0.95$), an instance of the
WWW ($\rho=0.5165$), two neural networks ($0.41\le\rho\le 0.44$),
two e-mail networks ($0.19\le\rho\le 0.23$), two word association
networks ($0.12\le\rho\le 0.19$) and 43 metabolic networks
($0.006\le\rho\le 0.052$). In particular, for the 53 snapshots of
the World Trade Web considered, the use of $\rho$ allows to properly
track the evolution of reciprocity over time, as we shall discuss in
Section \ref{sec_wtw}. The negatively correlated networks considered
are two shareholding networks ($-0.0034\le\rho\le -0.0012$) and 28
food webs ($-0.13\le\rho\le -0.01$). The case of minimum reciprocity
will be discussed in Section \ref{sec_minimum}.

The analysis reported above reveals that real networks display
nontrivial reciprocity patterns and are always either correlated or
anticorrelated. This result is very important, since theoretical
studies have shown that a nontrivial degree of reciprocity affects
the properties of various dynamical processes taking place on
directed networks, such as epidemic spreading \cite{recipr_newman},
percolation \cite{recipr_boguna}, and localization
\cite{recipr_vinko}. The effects of reciprocity are even more
interesting on scale-free networks, where even an infinitely small
fraction of bidirectional links was shown to give rise to a phase
transition characterized by the onset of a giant strongly connected
component \cite{recipr_boguna}.

\subsection{Minimum Reciprocity\label{sec_minimum}}
As we mentioned, in principle the allowed range of $\rho$ is
${-1\le\rho\le 1}$. However, from Table \ref{tab} we note that while
the most correlated directed network in the set considered displays
$\rho=0.95$, which is almost equal to the largest possible value,
the most anticorrelated one displays only $\rho=-0.13$, which is
quite far from the lower bound $\rho=-1$. Still, for most of the 30
antireciprocal networks reported in the table the number of
reciprocated links is zero ($r=0$) and therefore the value of $\rho$
is the minimum \linebreak possible \cite{myreciprocity}.

\begin{table}[h]
\begin{centering}
\begin{center}
\begin{tabular}{l  c}
\hline
~~~~~~~~~~~~Network & Range of $\rho$\\
\hline
\textbf{Perfectly reciprocal} & $\rho=1$\\

World Trade Web (53 webs)& $0.68\le\rho\le 0.95$\\

World Wide Web (1 web)& $\rho=0.5165$\\

Neural Networks (2 webs)&$0.41\le\rho\le 0.44$\\

Email Networks (2 webs)&$0.19\le\rho\le 0.23$\\

Word Networks (2 webs) &$0.12\le\rho\le 0.19$\\

Metabolic Networks (43 webs) &$0.006\le\rho\le 0.052$\\

\textbf{Areciprocal}& $\rho=0$\\

Shareholding Networks (2 webs)&$-0.0034\le\rho\le -0.0012$\\

Food Webs (28 webs)&$-0.13\le\rho\le -0.01$\\

\textbf{Perfectly antireciprocal}& $\rho=-1$\\
\hline
\end{tabular}
\caption{Empirical values of $\rho$ (in decreasing order), for the
133 real networks analysed in Reference \cite{myreciprocity}. The
values reported show the significant digits with respect to the
statistical errors.} \label{tab}
\end{center}
\end{centering}
\end{table}

\begin{figure}[h]
\begin{center}
\includegraphics[width=.6\textwidth]{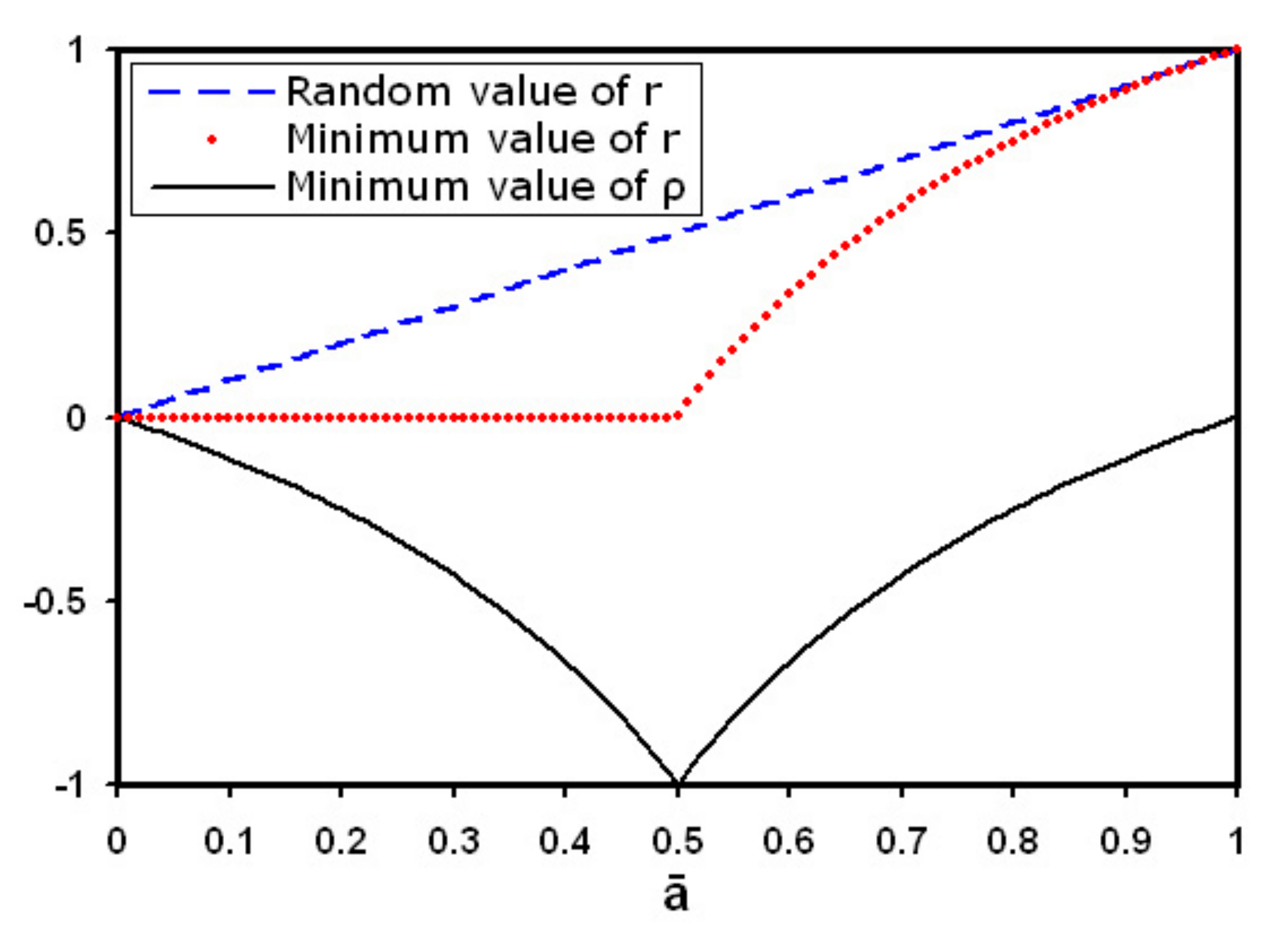}
\vspace{-1cm}
\end{center}
\caption{Behaviour of $r_{rand}$, $r_{min}$ and $\rho_{min}$ as functions of $\bar{a}$.\label{fig_rhomin}}
\end{figure}

This seemingly puzzling outcome can be explained as follows. Note
that Equation (\ref{rho}) implies that even in a network with $r=0$
the value of $\rho$ is always different from $-1$ unless
$\bar{a}=1/2$. This occurs because $\bar{a}=1/2$ is the only case
allowing perfect anticorrelation: in order to have $a_{ij}=1$
whenever $a_{ji}=0$, the adjacency matrix must be exactly
`half-filled' with unit entries, and the number of links must be
half the maximum possible one \cite{myreciprocity}. Remarkably, for
$\bar{a}\ne 1/2$ there are two different cases. In the `sparse'
range $\bar{a}<1/2$, the minimum value of $r$ is $r_{min}=0$ since
it is always possible to place all the links without having
reciprocal pairs. Consequently, Equation (\ref{rho}) implies that
${\rho_{min}=\bar{a}/(\bar{a}-1)}$. By contrast, in the `dense'
range $\bar{a}>1/2$ some links must be unavoidably placed between
the same pairs of vertices and therefore $r>0$. More precisely,
since the number of vertex pairs is ${N(N-1)/2}$, the minimum number
of reciprocal links is given by twice the number of links exceeding
this number, or in other words ${L^\lr_{min}=2[L-N(N-1)/2]}$.
Consequently, ${r_{min}=2-1/\bar{a}}$ and
${\rho_{min}=(\bar{a}-1)/\bar{a}}$. Putting these results together,
we have \be\label{rmin} r_{min}=\left\{\begin{array}{ll} 0 & \quad
\textrm{if $\bar{a}\le 1/2$}
\\&\\
2-\displaystyle\frac{1}{\bar{a}} & \quad \textrm{if $\bar{a}>1/2$}
\end{array}\right.
\ee
and
\be\label{rhomin}
\rho_{min}=\left\{\begin{array}{ll}
\displaystyle \frac{\bar{a}}{\bar{a}-1} & \quad \textrm{if $\bar{a}\le 1/2$}
\\&\\
\displaystyle \frac{\bar{a}-1}{\bar{a}} & \quad \textrm{if $\bar{a}>1/2$}
\end{array}\right.
\ee Both trends, together with the simple behaviour of
$r_{rand}=\bar{a}$ for an antireciprocal network, are shown as
functions of $\bar{a}$ in Figure \ref{fig_rhomin}.

\subsection{Related Topological Properties\label{sec_related}}
In this section we introduce various topological properties related
to the reciprocity of a network. We will refer again to Figure
\ref{fig_dirundgraph} to illustrate many of the properties discussed
in this section. The local quantities that characterize each vertex
$i$ are the \emph{in-degree} $k^{in}_i$ and the \emph{out-degree}
$k^{out}_i$, defined as the number of in-coming and out-going links
respectively: \bea
k^{in}_i&=&\sum_{j\ne i}a_{ji}\\
k^{out}_i&=&\sum_{j\ne i}a_{ij} \eea In the example shown in Figure
\ref{fig_dirundgraph}a, vertex $A$ has $k^{in}_A=4$ in-coming links
and $k^{out}_A=3$ out-going links. Unfortunately, these commonly
used quantities do not carry information about the reciprocity,
since they do not tell us if the in-coming and out-going links of a
vertex $i$ `overlap' completely, partly or not at all. As a way to
measure the overlap between the sets of in-coming and out-going
links of a vertex $i$, the \emph{reciprocated degree} $k^\lr_i$ was
defined
\cite{myreciprocity,mymultispecies,recipr_newman,recipr_boguna} as
the number of `reciprocal neighbours' (vertices joined by two
reciprocal links) of $i$: \be k^\lr_i\equiv\sum_{j\ne i}a_{ij}a_{ji}
\label{eq_kboth} \ee In the example shown in Figure
\ref{fig_dirundgraph}a, vertex $A$ has $k^{\lr}_A=2$ reciprocal
neighbours. As extreme examples, in a purely bidirectional network
($\rho=1$) there is complete overlap and
${k^\lr_i=k^{in}_i=k^{out}_i\ \forall i}$, while in a purely
unidirectional network ($\rho=\rho_{min}<0$) there is no overlap and
${k^\lr_i=0\ \forall i}$. One could think of $k^\lr_i$ as the result
of a kind of `attraction' or `repulsion' between the in-coming and
out-going links of vertex $i$, and of $\rho$ as an average strength
of the corresponding (positive or negative) interaction.

As we mentioned, the knowledge of $k^{in}_i$ and $k^{out}_i$ alone
is not enough to know $k^\lr_i$. It only informs us about the
maximum possible overlap, which is \be
(k^\lr_i)_{max}=\min\{k^{in}_i,k^{out}_i\} \ee In the case shown in
Figure \ref{fig_dirundgraph}a, $(k^\lr_A)_{max}=3$. If the total
number $N$ of vertices is known, then $k^{in}_i$ and $k^{out}_i$ can
also tell us about the minimum overlap, which is \be
(k^\lr_i)_{min}=\left\{\ba{ll}
0&\quad \textrm{if}\ k^{in}_i+k^{out}_i\le N-1\\
k^{in}_i+k^{out}_i-(N-1)&\quad \textrm{if}\ k^{in}_i+k^{out}_i> N-1
\ea\right. \ee depending on the possibility to place in-coming and
out-going links without overlap. The above expression is the
analogous of Equation (\ref{rmin}) for individual vertices. In the
case shown in Figure \ref{fig_dirundgraph}a, $(k^\lr_A)_{min}=2$.
Indeed, in the example considered it would be impossible to achieve
a value of $k^{\lr}_A$ smaller than that realised, given the values
of $k^{in}_A$ and $k^{out}_A$. It would also be impossible to
achieve a value of $k^{\lr}_A$ larger than $3$. In general, even the
joint knowledge of the in- and out-degrees $\{k^{in}_i\}$ and
$\{k^{out}_i\}$ of all vertices, or similarly the joint degree
distribution $P(k^{in}=k,k^{out}=k')$ that a randomly chosen vertex
has in-degree $k$ and out-degree $k'$, cannot characterize the
reciprocity of the network. What can be extracted from these
quantities is only the maximum and minimum numbers of reciprocated
links, an information analogous to that leading to Equation
(\ref{rhomin}).

By contrast, the \emph{three} degree sequences $\{k^{in}_i\}$,
$\{k^{out}_i\}$ and $\{k^\lr_i\}$ specify the connectivity
properties including the reciprocity. Summing over all vertices
gives the same information as Equations (\ref{L}) \linebreak and
(\ref{Lboth}), and $\rho$ can then be easily computed.
Alternatively, it is also possible to define the
\emph{non-reciprocated in-degree} $k^\gets_i$ and the
\emph{non-reciprocated out-degree} $k^\to_i$ of a vertex $i$ as the
number of in-coming and out-going links that are not reciprocated
respectively: \bea
k^\gets_i&\equiv&\sum_{j\ne i}a_{ji}(1-a_{ij})=k^{in}_i-k^\lr_i\\
k^\to_i&\equiv&\sum_{j\ne i}a_{ij}(1-a_{ji})=k^{out}_i-k^\lr_i \eea
In the example shown in Figure \ref{fig_dirundgraph}a, vertex $A$
has $k^\gets_A=2$ and $k^\to_A=1$. The information specified by the
three degree sequences $\{k^\gets_i\}$, $\{k^\to_i\}$ and
$\{k^\lr_i\}$ is the same as that carried by $\{k^{in}_i\}$,
$\{k^{out}_i\}$ and $\{k^\lr_i\}$. Note that the \emph{total degree}
$k^{tot}_i$ can be expressed in the equivalent forms \bea
k^{tot}_i&=&\sum_{j\ne i}(a_{ji}+a_{ij})\\
&=&k^{in}_i+k^{out}_i\nonumber\\
&=&k_i^\gets+k_i^\to+2k_i^\lr\nonumber
\eea

The above quantities also come into play whenever a directed graph
is regarded as an undirected one by simply ignoring the direction of
the links. We will consider this problem in a real-world case in
Section \ref{sec_wtw}. The undirected projection of a directed graph
is an undirected graph where each pair of vertices is connected by
an undirected edge if \emph{at least one} directed link
(irrespective of its direction) is present between them in the
original directed graph. Figure \ref{fig_dirundgraph}b reports the
undirected version of the directed graph of Figure
\ref{fig_dirundgraph}a. If $A$ is the adjacency matrix of the
original directed network, then the adjacency matrix $B$ of the
projected undirected network has entries \be\label{b}
b_{ij}=a_{ij}+a_{ji}-a_{ij}a_{ji} \ee and is now symmetric, as for
any undirected network. Each vertex $i$ in the undirected graph is
now simply characterized by its \emph{undirected degree} $k_i$:
\bea\label{k}
k_i&=&\sum_{j\ne i}b_{ji}\\
&=&k^{in}_i+k^{out}_i-k^\lr_i\nonumber\\
&=&k^\gets_i+k^\to_i+k^\lr_i\nonumber \eea The number of links $L^u$
in the undirected network is \be
L^u=\sum_{i<j}b_{ij}=\frac{1}{2}\sum_i
k_i=L-\frac{1}{2}L^\lr=\left(1-\frac{r}{2}\right)L \label{eq_Lu} \ee
and the link density, or connectance, of the undirected network is
the ratio between $L^u$ and the maximum number of undirected links,
{\em i.e.} \be
\bar{b}\equiv\frac{2\sum_{i<j}b_{ij}}{N(N-1)}=\frac{2L^u}{N(N-1)}=(2-r)\bar{a}
\label{eq_cu} \ee which is in an interesting relation with Equation
(\ref{eq_conn}). From the above equations, which can be checked
explicitly in the example shown in Figure \ref{fig_dirundgraph}, it
is clear that the knowledge of $k^{in}_i$ and $k^{out}_i$ is not
enough to determine $k_i$. Again, a crucial role is played by
$k^\lr_i$ and consequently by the reciprocity of the network. For
perfectly antireciprocal networks $k^\lr_i=0$ and $k_i=k^{tot}_i$,
while for perfectly reciprocal ones $k_i=k^\lr_i=k^{tot}/2$. More in
general, the knowledge of a directed topological property is not
enough to determine the corresponding property in the projected
undirected graph. The missing information is carried by the
reciprocity structure of the network.

In what follows, it will be useful to evaluate the expectation
values of the above quantities across various graph ensembles.
Therefore, before discussing specific cases, we briefly develop a
formalism useful in an ensemble setting. In a graph ensemble, each
link has an associated probability of \linebreak occurrence
\cite{symmetry1}. The information relevant to the reciprocity
structure is captured by two different probabilities. The first one
is the \emph{marginal} probability \be\label{pij} p_{ij}\equiv
p(i\to j)=\langle a_{ij}\rangle \ee that a directed link from $i$ to
$j$ is there, irrespective of the presence of the reciprocal link.
The second one is the \emph{conditional} probability $r_{ij}$ that a
directed link from vertex $i$ to vertex $j$ is there \emph{given
that} the reciprocal link from $j$ to $i$ is there: \be\label{rij}
r_{ij}\equiv p(i\to j|i\gets j) \ee The trivial case, where the
occurrence of reciprocal links is only due to chance, is when the
event $i\gets j$ has no influence on the event $i\to j$, so that
$r_{ij}$ is equal to the marginal probability $p_{ij}$. By contrast,
if $r_{ij}>p_{ij}$ ($r_{ij}<p_{ij}$), the presence of two mutual
links between $i$ and $j$ is more (less) likely than expected by
chance.

From the two probabilities above, a range of properties related to
the reciprocity structure can be derived. For instance, the
probability $p^\lr_{ij}$ that $i$ and $j$ are connected by two
reciprocal links is \be\label{plr} p^\lr_{ij}\equiv p(i\to j\cap
i\gets j)=\langle a_{ij}a_{ji}\rangle=r_{ij}p_{ji}=r_{ji}p_{ij} \ee
and the probability $p^\to_{ij}$ that a single link from $i$ to $j$
is there, with no reciprocal one from $j$ to $i$, is \be\label{pto}
p^\to_{ij}\equiv p(i\to j\cap i\nleftarrow j) =\langle
a_{ij}(1-a_{ji})\rangle=p_{ij}-p^\lr_{ij}=p_{ij}(1-r_{ji}) \ee
Consequently, the expected values of $k^{in}_i$, $k^{out}_i$ and
$k^\lr_i$ are
\begin{eqnarray}
\langle k^{in}_i\rangle&=&\sum_{j\ne i}p_{ji}\\
\langle k^{out}_i\rangle&=&\sum_{j\ne i}p_{ij}\\
\langle k^{\lr}_i\rangle&=&\sum_{j\ne i}r_{ij}p_{ji}
\end{eqnarray}
Similarly, the expectation value of the total number of directed links is
\be
\langle L\rangle =\sum_{i\ne j}\langle a_{ij}\rangle=\sum_{i\ne j}p_{ij}
\ee
and that of the number of reciprocated links is
\be
\langle L^\lr\rangle =\sum_{i\ne j}\langle a_{ij}a_{ji}\rangle=\sum_{i\ne j}p^\lr_{ij}=\sum_{i\ne j}r_{ij}p_{ji}
\ee
Therefore we can write down an expression for the expected value of $r$ across the ensemble:
\be
\langle r\rangle=\frac{\sum_{i\ne j}r_{ij}p_{ji}}{\sum_{i\ne j}p_{ij}}
\label{eq_rexp}
\ee
Similarly, the expected correlation coefficient $\rho$ can be expressed as
\be\label{rhor}
\langle \rho\rangle=
\frac{\sum_{i\ne j}p_{ij}r_{ji}-(\sum_{i\ne j}p_{ij})^2/N(N-1)}
{\sum_{i\ne j}p_{ij}-(\sum_{i\ne j}p_{ij})^2/N(N-1)}
\ee
The above relations will be useful later on.\\

It is also possible to exploit $p_{ij}$, $r_{ij}$ and $p_{ij}^\lr$
to obtain the probability that an undirected link $(i-j)$ exists
between vertices $i$ and $j$ in the undirected projection of the
graph defined by Equation (\ref{b}). If $q_{ij}\equiv p(i-j)$
denotes this \emph{undirected} connection probability, then Equation
(\ref{b}) implies \be q_{ij}\equiv p(i-j)=\langle
b_{ij}\rangle=p_{ij}+p_{ji}-p^\lr_{ij}=p_{ij}+p_{ji}-r_{ij}p_{ji}
\label{qij} \ee Therefore the expectation value of the undirected
degree $k_i$ defined in Equation (\ref{k}) is \bea\label{kexp}
\langle k_i\rangle&=&\sum_{j\ne i}q_{ji}\\
&=&\langle k^{in}_i\rangle+\langle k^{out}_i\rangle-\langle k^\lr_i\rangle\nonumber\\
&=&\langle k^\gets_i\rangle+\langle k^\to_i\rangle+\langle k^\lr_i\rangle\nonumber
\eea
Similarly, the expected number of undirected links is
\be
\langle L^u\rangle=\sum_{i<j}q_{ij}=\frac{1}{2}\sum_i \langle k_i\rangle=\langle L\rangle-\frac{1}{2}\langle L^\lr\rangle
\label{eq_Luexp}
\ee
and the expected undirected connectance is
\be
\langle\bar{b}\rangle=\frac{2\sum_{i<j}q_{ij}}{N(N-1)}=\frac{2\langle L^u\rangle}{N(N-1)}
\label{eq_cuexp}
\ee

\section{Reciprocity, Link Reversal Symmetry, and Ensemble Equiprobability\label{sec_relation}}
We can now discuss the relation between the reciprocity of networks
(\ref{sec_reciprocity}), ensemble equiprobability (Section
\ref{sec_ensembles}), and the two types of link reversal symmetry
defined in Section \ref{sec_reversal}, {\em i.e.} \emph{transpose
equivalence} and \emph{transpose equiprobability}. As we shall try
to highlight, different invariances are captured by different
topological properties, including the two measures of reciprocity we
have introduced. This shows that an in-depth understanding of graph
symmetries can indicate more informative definitions of topological
properties. We start by stressing again that if $r=1$, or
equivalently $\rho=1$, then every edge is reciprocated. This means
that the network has the first type of link reversal invariance,
{\em i.e.} transpose equivalence: the adjacency matrix $A$ is
symmetric ($A=A^T$). The quantities $r$ and $\rho$ are therefore
both informative with respect to transpose equivalence. By contrast,
as we now show they carry different pieces of information about
ensemble equiprobability and, as a particular case of it, the second
type of link reversal invariance, {\em i.e.} transpose
equiprobability. As we mentioned, both symmetries are related to an
ensemble of graphs rather than to a single network. We can therefore
exploit the expressions derived in Section \ref{sec_related} to
obtain the expected reciprocity structure in specific graph
ensembles. The natural class of graph ensembles relevant to this
problem is the directed version of the maximum-entropy models with
specified constraints \cite{symmetry1} that we have briefly recalled
in \linebreak Section \ref{sec_ensembles}. These ensembles provide
us with a null model against which, as we anticipated in Section
\ref{sec_link}, it is important to compare the empirically observed
reciprocity. For directed networks, (grand)canonical graph ensembles
consist of $2^{N(N-1)}$ possible directed graphs with no self-loops,
each described by a Hamiltonian $H(G)$ and by the corresponding
maximum-entropy probability \linebreak $P(G)=e^{-H(G)}/Z$.

As a first example, we consider the \emph{directed random graph},
which is the directed analogue of the model defined by Equation
(\ref{eq_Hrand}) and corresponds to the Hamiltonian \be H(G)=\theta
L(G)=\theta\sum_{i\ne j}a_{ij}(G) \label{eq_Hdirrand} \ee where now
$L(G)$ is the number of \emph{directed} links. In such a model, a
directed link from vertex $i$ to vertex $j$ is drawn with constant
probability $p\equiv e^{-\theta}/(1+e^{-\theta})$, independently of
all other links. That is, also the reciprocal link from $j$ to $i$
is drawn independently and with the same probability $p$. Due to the
statistical independence of reciprocal edges, in this model the
conditional probability $r_{ij}$ reduces to the marginal one
$p_{ij}$. Putting these results together, we have: \be
r_{ij}=p_{ij}=p=\frac{e^{-\theta}}{1+e^{-\theta}} \ee Inserting the
above relation into Equation (\ref{eq_rexp}) one finds that the
expected value of $r$ is \be \langle r\rangle=p \ee If, in analogy
with the undirected random graph discussed in Section
\ref{sec_ensembles} and according to the maximum likelihood
principle \cite{mylikelihood}, $p$ is set to the value
$p^*=L(G^*)/N(N-1)$ producing a null model of a real network $G^*$
with $L(G^*)$ directed links and connectance
$\bar{a}(G^*)=L(G^*)/N(N-1)$, then the expected value of $r$ in the
directed random graph model is \be r_{rand}= p^*=\bar{a}(G^*)
\label{eq_rrmodel} \ee Similarly, the expected value of $\rho$ under
the same null model is \be \rho_{rand}=\frac{r_{rand}- p^*}{1-
p^*}=0 \label{eq_rhorho} \ee The above results prove what we
anticipated in Equations (\ref{rand}) and (\ref{rhorand}). Note that
the directed random graph model defined by Equation
(\ref{eq_Hdirrand}) is symmetric under transpose equiprobability:
since $\theta$ is a global parameter, one has $H(A)=H(A^T)$ (where
$A$ denotes the adjacency matrix of graph $G$) irrespective of the
symmetry of the real network $G^*$. A consequence of this invariance
is that in the null model the expected in-degree and out-degree of
any vertex are equal: \be \langle k^{in}_i\rangle=\langle
k^{out}_i\rangle =p^*(N-1)\qquad \forall i \label{eq_kkrand} \ee
irrespective of whether they are equal in the real network.
Similarly, the expectation values of all other directed properties
are invariant under link reversal, {\em i.e.} exchanging the inward
and outward directions. We can also rephrase the differences between
$r$ and $\rho$ in terms of their performance with respect to
transpose equiprobability in the random graph model as follows. The
reciprocity measure $r$ is completely uninformative with respect to
transpose equiprobability, since its behaviour under even this
simple null model is not universal and depends on the link density
of the network. By contrast, $\rho$ is informative since the
transpose equiprobability of the directed random graph model
translates into a universal value $\rho_{rand}=0$.

Another case of interest is the \emph{directed configuration model},
defined by a generalisation of \linebreak Equation (\ref{eq_Hconf})
corresponding to the enforcement of both the in-degree and the
out-degree sequences $\{k^{in}_i\}$ and $\{k^{out}_i\}$ as
constraints: \be H(G)=\sum_i \left[ \theta_i^{in}
k^{in}_i(G)+\theta_i^{out}k^{out}_i(G)\right]= \sum_{i\ne
j}(\theta_i^{out}+\theta_j^{in})a_{ij}(G) \label{eq_Hdirconf} \ee In
this model, two reciprocal edges are again statistically
independent, therefore the conditional probability $r_{ij}$ equals
the marginal one $p_{ij}$, which is \be r_{ij}=p_{ij}=\frac{x_i
y_j}{1+x_i y_j} \label{eq_pijdirconf} \ee where $x_i\equiv
e^{-\theta^{out}_i}$ and $y_i\equiv e^{-\theta^{in}_i}$. The above
expression generalises Equation (\ref{eq_pij}) to directed graphs.
If the directed configuration model is used as a null model of a
real network $G^*$, a discussion similar to that leading to Equation
(\ref{eq_likelihoodk}) in the undirected case shows that the
parameter values $\{x^*_i\}$ and $\{y^*_i\}$ indicated by the
maximum likelihood principle are those satisfying the $2N$ coupled
equations
\begin{eqnarray}
\langle k^{in}_i\rangle&=&\sum_{j\ne i}\frac{x^*_j y^*_i}{1+x^*_j y^*_i}=k^{in}_i(G^*)\qquad \forall i
\label{eq_likelihoodkin}\\
\langle k^{out}_i\rangle&=&\sum_{j\ne i}\frac{x^*_i y^*_j}{1+x^*_i y^*_j}=k^{out}_i(G^*)\qquad \forall i
\label{eq_likelihoodkout}
\end{eqnarray}
ensuring that both the expected in-degree and out-degree sequences
equal the empirical ones. Note that, unlike the directed random
graph, in this model the transpose equiprobability symmetry does not
hold: Equation (\ref{eq_Hdirconf}) implies that in general $H(A)\ne
H(A^T)$. Only if $\theta^{in}_i=\theta^{out}_i$, or equivalently
$x_i=y_i$, then $H(A)=H(A^T)$. From Equations
(\ref{eq_likelihoodkin}) and (\ref{eq_likelihoodkout}) we see that
this only occurs \linebreak if $k^{in}_i(G^*)= k^{out}_i(G^*)$ $\forall i$,
{\em i.e.} if in the real network the in-degree and the out-degree
of all vertices are equal. In such a case, in analogy with Equation
(\ref{eq_kkrand}), one has that the inward and outward expected
topological properties in the null model are equal, and the
transpose equiprobability symmetry holds. However, if
$k^{in}_i(G^*)\ne k^{out}_i(G^*)$ for some $i$, the transpose
equiprobability symmetry does not hold.

\begin{figure}[h]
\begin{center}
\includegraphics[width=0.5\textwidth]{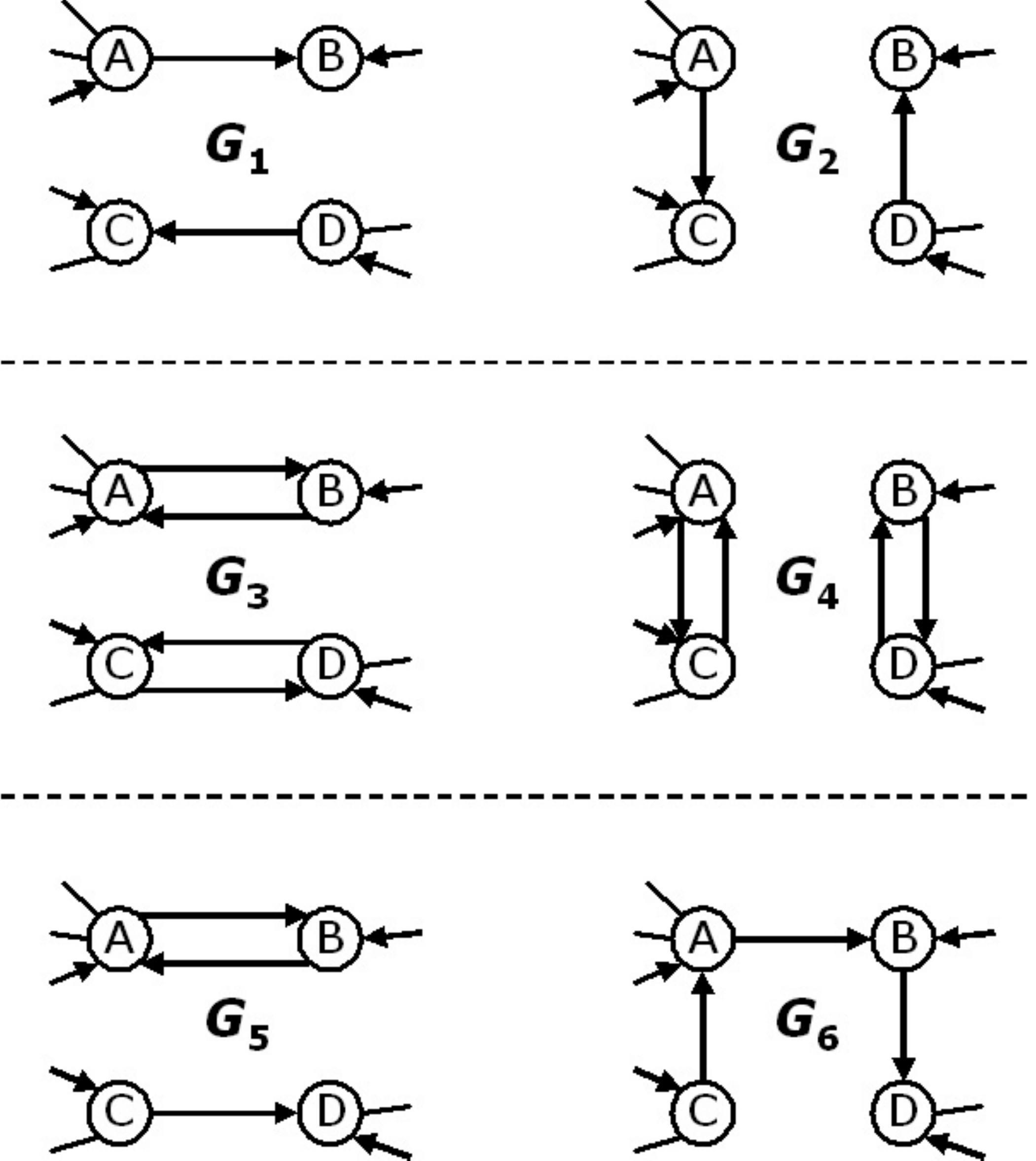}
\end{center}
\caption{In the directed version of the configuration model, the local rewiring algorithm \cite{symmetry1} has various generalizations. If one requires that only the two degree sequences $\{k^{in}_i\}$ and $\{k^{out}_i\}$ are preserved, with $H(G)$ defined by Equation (\ref{eq_Hdirconf}), then each graph on the left has the same probability of occurrence as the corresponding graph on the right, since $H(G_1)=H(G_2)$, $H(G_3)=H(G_4)$ and $H(G_5)=H(G_6)$. By additionally requiring that also $\{k^\lr_i\}$ is preserved, and redefining $H(G)$ as in Equation (\ref{eq_Hrecconf}), the above symmetry of the Hamiltonian is broken: $G_5$ and $G_6$ are no longer equiprobable since now $H(G_5)\ne H(G_6)$.
\label{fig_equidirgraphs}}
\end{figure}

In the directed configuration model, all the graphs with the same
in- and out-degree sequences are equiprobable, irrespective of the
number of mutual links arising in them. This produces a trivial
reciprocity structure.  As an example, consider Figure
\ref{fig_equidirgraphs}, which portrays various directed
generalisations of the \emph{local rewiring algorithm} introduced
for undirected networks \cite{symmetry1}. If $H(G)$ is defined by
\linebreak Equation (\ref{eq_Hdirconf}), then each graph on the left
has the same probability of occurrence as the corresponding graph on
the right, since $H(G_1)=H(G_2)$, $H(G_3)=H(G_4)$ and
$H(G_5)=H(G_6)$. However, while the two graphs $G_1$ and $G_2$, and
similarly the two graphs $G_3$ and $G_4$, have the same reciprocity,
the graphs $G_5$ and $G_6$ have different reciprocity, even if they
occur with the same probability in the ensemble defined by the
model. This means that the reciprocity structure of the network is
not preserved across the ensemble, just like any other property
except the in- and out-degree sequences, as required by the model.
This result confirms our discussion in Section \ref{sec_related},
where we showed that the two degree sequences $\{k^{in}_i\}$ and
$\{k^{out}_i\}$ alone do not specify the reciprocity of the network.
In analogy with the discussion leading to Equations
(\ref{eq_rrmodel}) and (\ref{eq_rhorho}) for the directed random
graph model, it is possible to study the reciprocity generated by
chance in the configuration model as the result of specifying given
degree \linebreak distributions \cite{vinko}. Conversely, it is also
possible to study the different problem of the influence of
reciprocal links on the degree distribution and degree correlations
\cite{vinko2}.

In order to generate an ensemble with nontrivial reciprocity, one
needs to enforce an additional constraint in the Hamiltonian. One
quite general possibility \cite{mymultispecies} is, according to our
discussion in Section \ref{sec_related}, to specify the \emph{three}
degree sequences $\{k^{in}_i\}$, $\{k^{out}_i\}$ and $\{k^\lr_i\}$:
\begin{eqnarray}
H(G)&=&\sum_i \left[ \theta_i^{in} k^{in}_i(G)+\theta_i^{out}k^{out}_i(G)+
\theta_i^{\lr}k^{\lr}_i(G)\right]\nonumber\\
&=&
\sum_{i\ne j}(\theta_i^{out}+\theta_j^{in})a_{ij}(G)+\sum_{i< j}(\theta_i^{\lr}+\theta_j^{\lr})a_{ij}(G)a_{ji}(G)
\label{eq_Hrecconf}
\end{eqnarray}
In the example shown in Figure \ref{fig_equidirgraphs}, in this model we still have $H(G_1)=H(G_2)$ and $H(G_3)=H(G_4)$, but now $H(G_5)\ne H(G_6)$ since the reciprocal degree sequence $\{k^\lr_i\}$ of the graphs $G_5$ and $G_6$ is different. So the addition of the extra term breaks the previous ensemble equiprobability symmetry of the Hamiltonian and restricts it to smaller equivalence classes. This implies that now the conditional and marginal connection probabilities are different: if we define $x_i\equiv e^{-\theta^{out}_i}$, $y_i\equiv e^{-\theta^{in}_i}$ and $z_i\equiv e^{-\theta^\lr_i}$ it can be shown \cite{mymultispecies} that
\begin{eqnarray}
p^\to_{ij}&=&\frac{x_i y_j}
{1+x_i y_j+x_j y_i+ x_i x_j y_i y_j z_i z_j}\label{eq_ptoij}\\
p^\lr_{ij}&=&\frac{x_i x_j y_i y_j z_i z_j }
{1+x_i y_j+x_j y_i+x_i x_j y_i y_j z_i z_j }\label{eq_plrij}
\end{eqnarray}
so that
\begin{eqnarray}
p_{ij}&=&p^\to_{ij}+p^\lr_{ij}=
\frac{x_i y_j+ x_i x_j y_i y_j z_i z_j}
{1+x_i y_j+x_j y_i+ x_i x_j y_i y_j z_i z_j}\label{eq_pij2}\\
r_{ij}&=&\frac{p^\lr_{ij}}{p_{ji}}=\frac{ x_i x_j y_i y_j z_i z_j}
{x_i y_j+x_i x_j y_i y_j z_i z_j }=
\frac{  x_j y_i z_i z_j}{1+ x_j y_i z_i z_j }\label{eq_rij2}
\end{eqnarray}

In this case the maximum likelihood principle \cite{mylikelihood}
indicates that, in order to provide a null model of a real network
$G^*$, the parameters $\{x_i\}$, $\{y_i\}$ and $\{z_i\}$ must be set
to the particular values $\{x^*_i\}$, $\{y^*_i\}$ and $\{z^*_i\}$
satisfying the $3N$ coupled equations
\begin{eqnarray}
\langle k^{in}_i\rangle&=&\sum_{j\ne i}\frac{x^*_i y^*_j+ x^*_i x^*_j y^*_i y^*_j z^*_i z^*_j}
{1+x^*_i y^*_j+x^*_j y^*_i+ x^*_i x^*_j y^*_i y^*_j z^*_i z^*_j}=k^{in}_i(G^*)\qquad \forall i
\label{eq_likelihoodkinrec}\\
\langle k^{out}_i\rangle&=&\sum_{j\ne i}\frac{x^*_j y^*_i+ x^*_i x^*_j y^*_i y^*_j z^*_i z^*_j}
{1+x^*_i y^*_j+x^*_j y^*_i+ x^*_i x^*_j y^*_i y^*_j z^*_i z^*_j}=k^{out}_i(G^*)\qquad \forall i\label{eq_likelihoodkoutrec}\\
\langle k^{\lr}_i\rangle&=&\sum_{j\ne i}\frac{x^*_i x^*_j y^*_i y^*_j z^*_i z^*_j}
{1+x^*_i y^*_j+x^*_j y^*_i+ x^*_i x^*_j y^*_i y^*_j z^*_i z^*_j}=k^{\lr}_i(G^*)\qquad \forall i
\label{eq_likelihoodkbothrec}
\end{eqnarray}
ensuring that the expectation values of the three degree sequences
equal the empirical ones. Again, we see that in this model the
transpose equiprobability symmetry only holds if the real network
$G^*$ has $k^{in}_i(G^*)=k^{out}_i(G^*)$ $\forall i$. In such a
case, from the above equations one finds $x^*_i=y^*_i$ $\forall i$
which also implies $p_{ij}=p_{ji}$ and $H(A)=H(A^T)$ so that all the
expected topological properties have inward/outward invariance.
Otherwise, the symmetry does not hold. A particular case of the
above model turns out to empirically describe the World Trade Web,
as we discuss in the next section.

\section{Symmetries, Symmetry Breaking and the Evolution of World Trade\label{sec_wtw}}
We now present an important real-world application of the concepts
introduced so far, {\em i.e.} the evolution of the international
trade network. The World Trade Web (WTW in the following) is the
global network of import/export trade relationships among all world
countries
\cite{serrano,mywtw,myalessandria,myinterplay,5,giorgiowtw}. We
already encountered the WTW in Section \ref{sec_rho} among the other
networks reported in Table \ref{tab}. In the WTW, a vertex
represents one country and a directed link represents the existence
(during the period considered, usually one year) of an export
relationship from one country to another country. The WTW is in
principle a weighted network, since trade intensities can be
measured by their (highly heterogeneous) total monetary values
aggregated over the period. Therefore the properties of the network
can be measured on a weighted basis \cite{5,giorgiowtw}. However,
here we will consider the WTW as a binary network, and only refer to
its purely topological properties. As we will show, even this simple
picture is extremely interesting and allows an informative study of
the international trade system. In particular, we will study how the
network has evolved in time starting from the year 1950, and how a
joint analysis of the trends displayed by different topological
properties inform us about the change in the underlying symmetries.
If the WTW is regarded as an undirected graph, its structural
properties are remarkably stable over time, and indicate that the
network displays a clear invariance under transformations that
preserve its degree sequence. On the other hand, when the
directionality of trade is taken into account, the above symmetry is
broken and the intensity of this symmetry breaking changes in time.
A strong increase in reciprocity is observed, clearly evidencing
that a major structural change started taking place from the late
1970's onwards. The symmetry concepts developed in Reference
\cite{symmetry1} and in the previous sections will be employed to
suggest, or rule out, possible explanations for the observed
evolution of the WTW. In particular, we identify as candidate
explanations a strong embedding in economic space and a spatial
symmetry breaking in the production system, which is known to have
occurred starting from the late 1970's \cite{7,8} and could
therefore explain the simultaneous change in the reciprocity of the
network. Surprisingly, other mechanisms such as the increase in the
number of trade relationships, size effects and the formation of
trade agreements are not enough in order to explain the observed
evolution of the symmetry properties considered. This analysis
highlights the importance of identifying the behaviour of complex
systems under different types of symmetries, and of introducing
suitable measures that succeed in distinguishing between the latter.

\subsection{Undirected Symmetries\label{sec_undirected}}
\begin{figure}[h]
\begin{center}
\includegraphics[width=.65\textwidth]{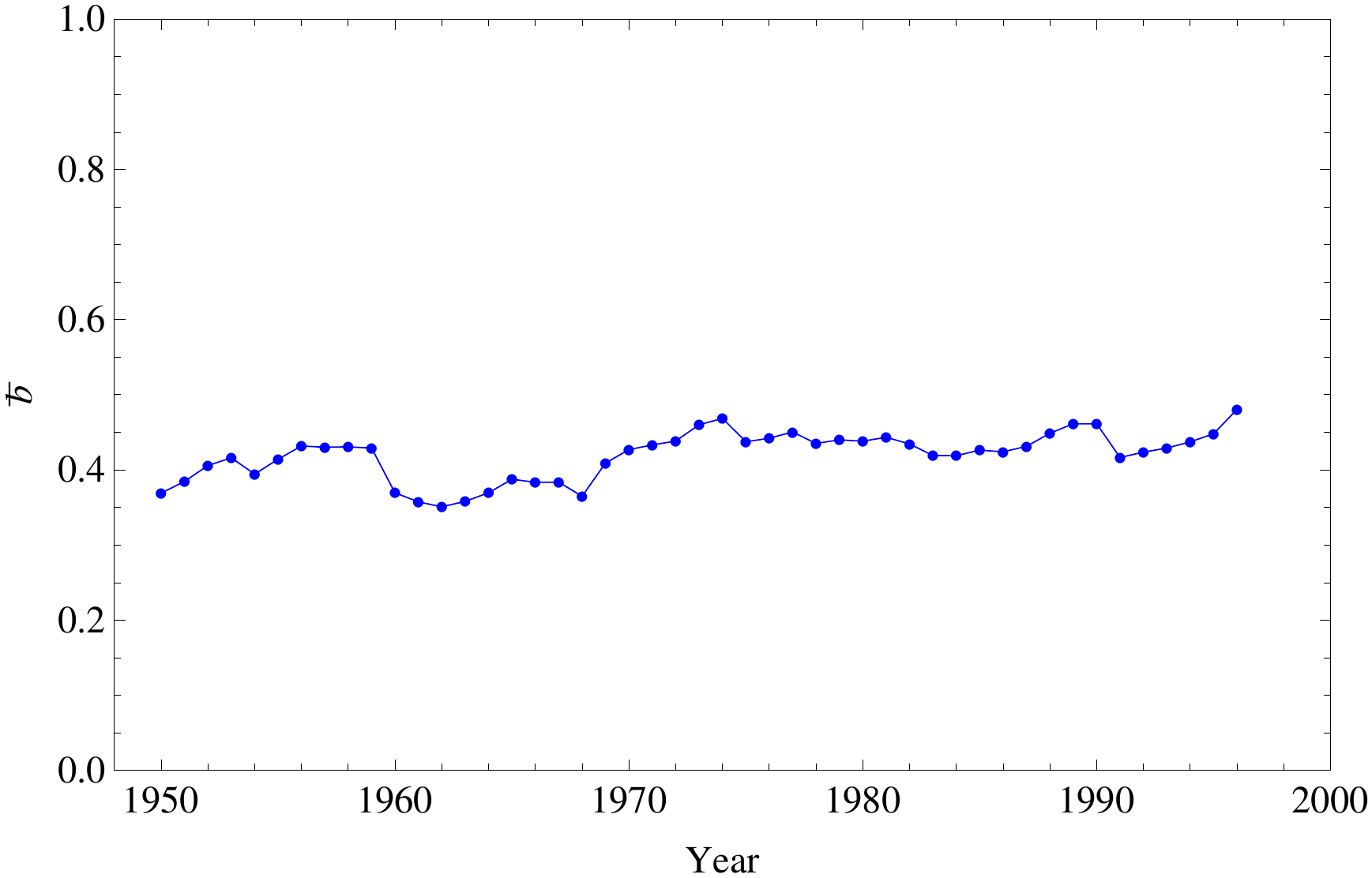}
\vspace{-0.5cm}
\end{center}
\caption{Evolution of the density $\bar{b}(t)$ of the undirected
version of the World Trade Web.\label{fig_density}}
\end{figure}

Various empirical results describing the topology of the WTW can be
combined in order to have a detailed picture of the underlying
symmetries. In this section we consider the undirected projection of
the network as defined in Section \ref{sec_related}, while in the
next one we consider the WTW as a directed network. A first
interesting observation, that will be useful in the following, is
that the undirected connectance $\bar{b}$ defined by Equation
(\ref{eq_cu}) remains almost constant during the time interval
considered, as shown in Figure \ref{fig_density}. This happens
despite the fact that the number $N(t)$ of world countries increases
significantly, due to a number of new independent states being
formed between 1948 and 2000.

Importantly, the constancy of the connectance does not mean that the
latter characterises the WTW satisfactorily. If we use the random
graph model as a null model of the WTW, the undirected connection
probability defined in Equation (\ref{qij}) is uniform:
$q_{ij}(t)=q(t)$. The maximum likelihood principle, in accordance
with Equation (\ref{eq_likelihoodL}), indicates the following choice
for this probability: \be q^*(t)=\frac{2L^u(t)}{N(t)[N(t)-1]}
\label{eq_qWTW} \ee However, the above choice generates trivial
expectations which are not in accordance with the empirical results,
in particular a Binomial degree distribution, a constant
(uncorrelated with the degree) average nearest neighbour degree and
a constant clustering coefficient. This means that the ensemble
equiprobability invariance of the random graph model under
transformations preserving the total number of links is not a
stochastic symmetry of the WTW in the sense explained in Section
\ref{sec_ensembles}.

By contrast, an important finding \cite{myrandomization,mywtw} is
that, in every snapshot of the network within the time window
considered, the undirected projection of the WTW is always
remarkably well reproduced by the  configuration model. This means
that, according to Equation (\ref{eq_pij}), the probability that a
trade relationship exists (irrespective of its direction) between
two countries $i$ and $j$ in a given year $t$ is \be
q_{ij}(t)=\frac{w^*_i(t) w^*_j(t)}{1+w^*_i(t) w^*_j(t)} \label{eq_q}
\ee where the parameters $\{w^*_i(t)\}$ are the solution of the $N$
coupled equations \be \langle k_i(t)\rangle =\sum_{j\ne
i}\frac{w^*_i (t)w^*_j(t)}{1+w^*_i(t) w^*_j(t)}=k_i(t)\qquad \forall
i \label{eq_likelihoodw} \ee which are equivalent to Equation
(\ref{eq_likelihoodk}). The accordance between the configuration
model and the real undirected WTW has been checked by studying
various higher-order properties, including the average nearest
neighbour degree and the clustering coefficient of all vertices, and
confirming that they are excellently reproduced by the model
\cite{myrandomization,mywtw}. The undirected WTW is therefore a good
example of a network whose higher-order properties can be traced
back to low-level constraints. According to our discussion in
Section \ref{sec_ensembles}, this  implies that the ensemble
equiprobability invariance displayed by the configuration model
under transformations preserving the degree sequence
\cite{symmetry1} is a stochastic symmetry of the real WTW. In turn,
this implies that in every snapshot of the WTW all vertices with the
same degree $k$ are statistically equivalent \cite{symmetry1}. That
is, the overall symmetry of the network under permutations of vertex
labels is broken down to distinct universality classes consisting of
vertices with the same degree. This is evident from the fact that,
in passing from the random graph model (where all vertices are
statistically equivalent) to the configuration model (where all
vertices with the same degree are statistically equivalent), the
connection probability changes from Equations (\ref{eq_qWTW}) to (\ref{eq_q}) and therefore acquires a dependence on the
variables $w_i^*$ and $w_j^*$, which in turn depend on the degree
sequence through Equation (\ref{eq_likelihoodw}). Unlike $q(t)$, the
probability $q_{ij}(t)$ is not uniform across all pairs of vertices,
but only across pairs of vertices with the same pair of degrees
$k_i$ and $k_j$. As shown in Equation (\ref{eq_cuexp}), the
following relation holds between the expected connectance
$\langle\bar{b}\rangle$ and the probability $q_{ij}(t)$: \be
\langle\bar{b}(t)\rangle=\frac{2\sum_{i<j}q_{ij}(t)}{N(N-1)}=\frac{2}{N(N-1)}\sum_{i<j}\frac{w^*_i
(t)w^*_j(t)}{1+w^*_i(t) w^*_j(t)} \ee Therefore the observed
stationarity of $\bar{b}$ shown in Figure \ref{fig_density}
indicates that, despite $q_{ij}(t)$ varies greatly among pairs of
world countries and also over time, its average across all pairs of
countries remains remarkably stable.

The accordance between the undirected WTW and the configuration
model means that the degree sequence is extremely informative, since
its knowledge allows one to obtain correct expectations about all
other topological properties. This implies that, in order to explain
the WTW topology at an undirected level, it is enough to explain its
degree sequence. Thus reproducing the degree sequence should be the
target of any model of the WTW topology, an important point that we
will discuss in Section \ref{sec_economic}. Whatever the cause of
the empirical degree sequence of WTW, this cause is the
symmetry-breaking phenomenon restricting the invariance of the
network to degree-preserving transformations.

\subsection{Directed Symmetries\label{sec_directed}}
We now come to the description of the WTW as a directed network,
which involves additional information. Note that, since the
configuration model reproduces the real WTW topology, and since in
this model different pairs of vertices are statistically
independent, then also the directed version of the model must be
reproduced by a model with independent pairs of vertices. What
remains to be clarified is whether the possible events that can
occur within a single pair of vertices are also statistically
independent, {\em i.e.} whether the conditional connection
probability $r_{ij}$ and the marginal connection probability
$p_{ij}$ defined in Section \ref{sec_related} are equal. In other
words, we need to characterise the reciprocity structure of the
network.

\begin{figure}[h]
\begin{center}
\includegraphics[width=.7\textwidth]{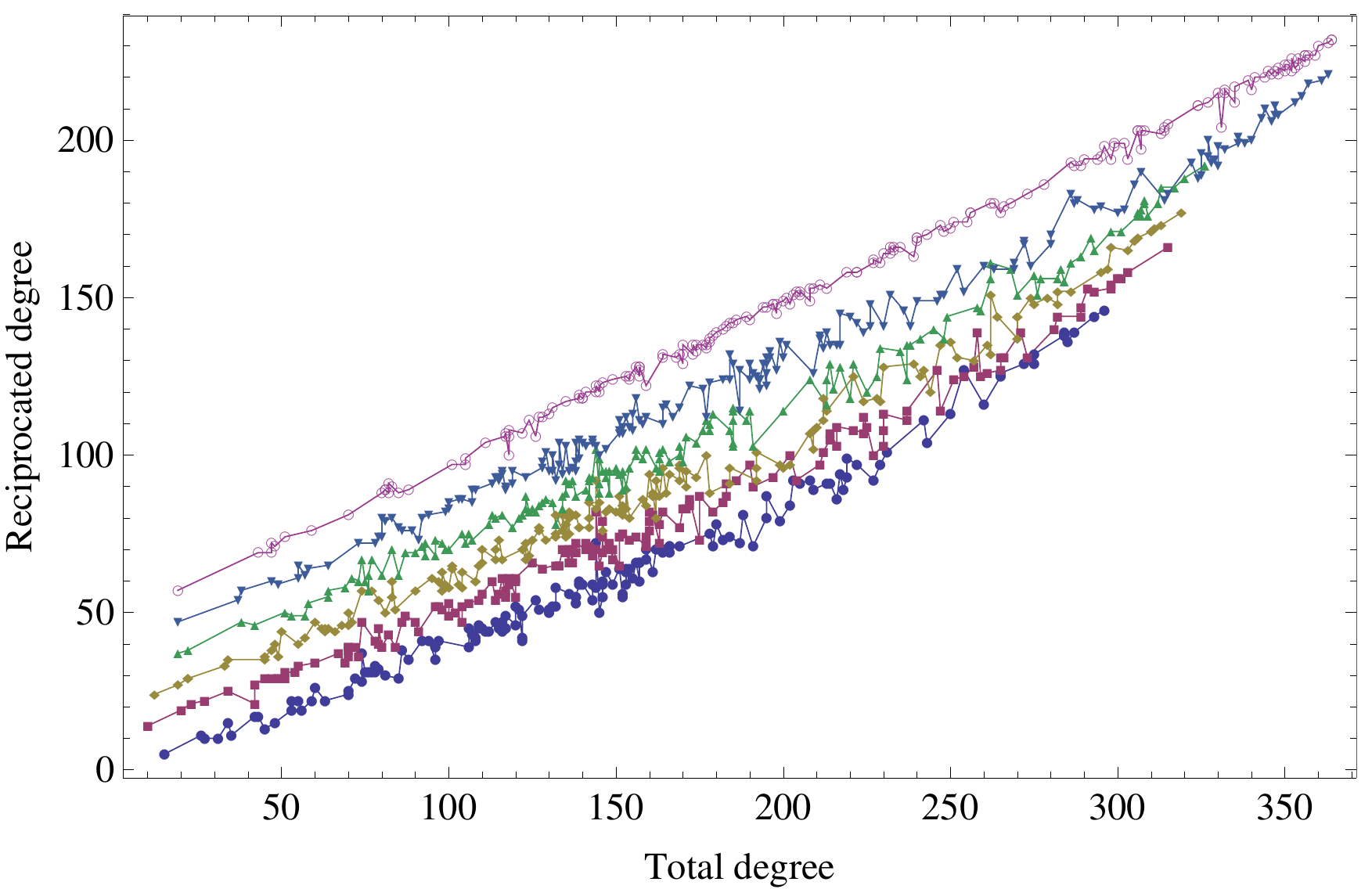}
\end{center}
\caption{Dependence of the reciprocated degree $k^{\leftrightarrow}_i$ on the total degree $k^{tot}_i=k^{in}_i+k^{out}_i$ in various snapshots of the WTW (from
bottom to top: $t=1975, 1980, 1985, 1990, 1995, 2000$). The
different curves have been shifted vertically for better visibility.
\label{fig_kboth}}
\end{figure}

To this end, a first useful result is that, irrespective of the year
$t$ considered, the in-degree and the out-degree of every vertex are
empirically found to be approximately equal
\cite{myreciprocity,myalessandria,myinterplay}, {\em i.e.} \be
k^{in}_i(t)\approx k^{out}_i (t)\qquad\forall i \label{eq_kin=kout}
\ee A second empirical result is that the reciprocated degree
$k^\lr_i(t)$ defined in Equation (\ref{eq_kboth}) is always
proportional to the total degree $k^{tot}_i(t)=k^{in}_i(t)+k^{out}_i
(t)$, with a time-dependent proportionality coefficient
\cite{myreciprocity,myalessandria}: \be k^{\lr}_i(t)\propto
k^{tot}_i (t)\qquad\forall i \label{eq_kbothktot} \ee This result is
shown in Figure \ref{fig_kboth} for various years $t$.

Taken together, these two results inform us about the structure of
the connection probabilities $p_{ij}$, $r_{ij}$ and $p_{ij}^\lr$
introduced in Section \ref{sec_related}. Indeed, since $\langle
k^{in}_i(t) \rangle=\sum_{j\ne i}p_{ji}(t)$ and $\langle
k^{out}_i(t) \rangle=\sum_{j\ne i}p_{ij}(t)$, the result in Equation
(\ref{eq_kin=kout}) can be rephrased as \be p_{ij}(t)\approx
p_{ji}(t)\qquad r_{ij}(t)\approx r_{ji}(t) \ee Similarly, since
$\langle k^\lr_i(t) \rangle=\sum_{j\ne i}r_{ij}(t)p_{ji}(t)$,
Equation (\ref{eq_kbothktot}) implies that $r_{ij}(t)$ is
independent of $i$ and $j$, {\em i.e.} the conditional connection
probability is uniform: \be r_{ij}(t)\approx r_0(t) \label{eq_rij0}
\ee The latter determines the proportionality coefficient relating
the reciprocated degree to the total degree as in Equation
(\ref{eq_kbothktot}): \be \langle k_i^\lr(t)\rangle=r_0(t)\sum_{j\ne
i}p_{ij}(t)=r_0(t)\langle
k^{out}_i(t)\rangle=\frac{r_0(t)}{2}\langle k^{tot}_i(t)\rangle
\label{eq_kr0} \ee Moreover, Equation (\ref{eq_rexp}) implies that
the expected reciprocity of the network at time $t$ coincides with
the conditional connection probability: \be \langle r(t)\rangle
\approx r_0(t) \label{eq_rr} \ee This result can be confirmed
independently,by measuring the observed reciprocity $r(t)$ and
checking that it is indeed approximately equal to the
proportionality coefficient $r_0(t)$ relating the reciprocated
degree to the total degree as in Equation (\ref{eq_kr0}), obtained
from a linear fit of the trends shown in Figure \ref{fig_kboth}
\cite{myreciprocity}. We will show more empirical results about the
reciprocity in Section \ref{sec_revolution}.

The uniformity of $r_{ij}(t)$ implies that the marginal connection
probability must be different from the conditional one. Otherwise,
the WTW would be well reproduced by the directed random graph model
introduced in Section \ref{sec_relation}, with $p(t)\approx r_0(t)$.
This possibility is ruled out by the fact that the two approximate
equalities $p_{ij}(t)\approx p_{ji}(t)$ and $r_{ij}(t)\approx
r_0(t)$, if inserted into Equation (\ref{qij}), imply \be
q_{ij}(t)\approx [2-r_0(t)]p_{ij}(t) \label{eq_qpWTW} \ee which,
together with Equation (\ref{eq_q}), implies that the WTW is well
described by the marginal connection probability \be
p_{ij}(t)\approx
\frac{1}{2-r_0(t)}q_{ij}(t)=\frac{1}{2-r_0(t)}\frac{w^*_i(t)
w^*_j(t)}{1+w^*_i(t) w^*_j(t)} \label{eq_pqWTW} \ee The above
marginal probability is not uniform as the directed random graph
model would predict, and is necessarily different from the uniform
conditional connection probability $r_0(t)$. This means that the
reciprocity of the WTW, whatever the snapshot considered, is
nontrivial. As another consequence, this result implies that a good
model of the WTW is not even provided by the directed configuration
model defined by Equation (\ref{eq_Hdirconf}), because the latter
predicts $r_{ij}= p_{ij}$ as shown in Equation
(\ref{eq_pijdirconf}). Therefore the directed representation of the
WTW does not display, as a stochastic symmetry, the ensemble
equiprobability invariance under transformations that preserve the
degree sequences $\{k^{in}_i\}$ and $\{k^{out}_i\}$.

As discussed in Section \ref{sec_relation}, a step forward the
simple directed configuration model is provided by the model defined
by Equation (\ref{eq_Hrecconf}), {\em i.e.} a maximum-entropy
ensemble of graphs with constraints given by the three degree
sequences $\{k^{out}_i\}$, $\{k^{in}_i\}$,  $\{k^{\lr}_i\}$
controlled by the Lagrange multipliers $\{\theta^{out}_i\}$,
$\{\theta^{in}_i\}$, $\{\theta^{\lr}_i\}$ or equivalently $\{x_i\}$,
$\{y_i\}$, $\{z_i\}$. We now prove various theoretical relations
describing what is implied when a uniform conditional connection
probability $r_{ij}=r_0$ is assumed as a further ingredient of this
model, and show that these relations are in excellent agreement with
all the empirical properties of the WTW discussed above, and
reconcile the undirected picture with the directed one. For brevity,
in our notation we drop the dependence of the various quantities on
the time $t$. First, note that, due to the equality
$p_{ij}r_{ji}=p_{ji}r_{ij}$ appearing for instance in Equation
(\ref{plr}), $r_{ij}=r_0$ implies \be p_{ij}=p_{ji} \ee and
automatically predicts both $\langle k^{in}_i\rangle=\langle
k^{out}_i\rangle$ and $\langle k^{\lr}_i\rangle=(r_0/2)\langle
k^{tot}_i\rangle$. In other words, the constancy of $r_{ij}$ implies
the symmetry of $p_{ij}$ and is enough to simultaneously explain the
two empirical properties of the WTW reported in Equations
(\ref{eq_kin=kout}) and (\ref{eq_kbothktot}). As another
consequence, one has $x_i=y_i$ $\forall i$ in \linebreak Equation
(\ref{eq_pij2}) so that Equation (\ref{eq_rij2}) becomes \be r_{ij}=
\frac{ x_i x_j z_i z_j}{1+ x_i x_j z_i z_j } \ee But under our
hypothesis the above expression must be a constant $r_0$, which is
only possible \linebreak if $x_i z_i=y_i z_i=\alpha$ where $\alpha$ is a
constant. This implies \be x_i=y_i=\alpha
z_i^{-1}\qquad\Leftrightarrow\qquad
\theta^{out}_i=\theta^{in}_i=-\ln\alpha-\theta^\lr_i
\label{eq_condition} \ee Therefore among the three parameters $x_i$,
$y_i$, $z_i$ there is only an independent one (say $x_i$). This
allows us to rewrite Equations (\ref{eq_ptoij}) and (\ref{eq_plrij})
as
\begin{eqnarray}
p^\to_{ij}&=&\frac{x_i x_j}
{1+(2+\alpha^2) x_i x_j }\label{eq_ptoijx}\\
p^\lr_{ij}&=&\frac{\alpha^2 x_i x_j }
{1+(2+\alpha^2) x_i x_j }\label{eq_plrijx}
\end{eqnarray}
and Equations (\ref{eq_pij2}) and (\ref{eq_rij2}) as
\begin{eqnarray}
p_{ij}&=&\frac{(1+\alpha^2)x_i x_j}
{1+(2+\alpha^2) x_i x_j }\label{eq_pijWTWrec}\\
r_{ij}&=&r_0=
\frac{  \alpha^2}{1+\alpha^2}\label{eq_pervert}
\end{eqnarray}
The last equation, if inserted into Equation (\ref{eq_rexp}),
implies \be \langle r\rangle=r_{ij}=\frac{  \alpha^2}{1+\alpha^2}
\label{eq_rexp2} \ee which clearly shows that in this model the
expected reciprocity $r$ coincides with the conditional probability
$r_{ij}$ and is uniquely determined by $\alpha$. Thus all the above
quantities could be expressed as functions of $r_0$ (or $\langle
r\rangle$) rather than $\alpha$, by exploiting the inverse relation
\be \alpha=\sqrt{\frac{r_0}{1-r_0}}=\sqrt{\frac{\langle
r\rangle}{1-\langle r\rangle}} \label{eq_invert} \ee Equation
(\ref{qij}) implies that under the above model the connection
probability in the undirected projection of the network is \be
q_{ij}=p_{ij}+p_{ji}-r_{ij}p_{ji}=\frac{(2+\alpha^2)x_i x_j}
{1+(2+\alpha^2) x_i x_j } \label{eq_qijWTWrec} \ee Together with
Equations (\ref{eq_pijWTWrec}) and (\ref{eq_invert}), the above
relation implies \be
q_{ij}=\frac{2+\alpha^2}{1+\alpha^2}p_{ij}=(2-r_0)p_{ij} \ee which
exactly reproduces the empirical property of the WTW shown in
Equation (\ref{eq_qpWTW}). Note that, if the parameters $\alpha$ and
$\{x_i\}$ are tuned to the values $\alpha^*$ and $\{x^*_i\}$ fitting
the model to the real network, \linebreak Equation
(\ref{eq_qijWTWrec}) coincides with Equation (\ref{eq_q}) once
$\alpha^*$ is reabsorbed into $w^*_i$ as follows: \be
w_i^*=x_i^*\sqrt{2+(\alpha^*)^2} \label{eq_reabsorb} \ee That is,
once the value of $\alpha^*$ enforcing the observed value of the
reciprocity is fixed, the values of $\{x_i^*\}$ determine the
undirected degree sequence exactly as $\{w_i^*\}$ in the undirected
configuration model. This important result indicates that the
undirected version of the directed model considered here coincides
with the undirected configuration model, and thus reconciles the
directed and undirected descriptions. Note that this is not true in
general: for instance, the undirected version of the directed
configuration model defined by Equation (\ref{eq_Hdirconf}) does
\emph{not} coincide with the undirected configuration model. It is
the nontrivial structure of the reciprocity of the WTW, manifest in
the uniformity of $r_{ij}$, that ensures this property. This result
can be confirmed by noticing that Equation (\ref{eq_condition})
implies that the Hamiltonian of the model, which in general has the
form in Equation (\ref{eq_Hrecconf}), in this case becomes
\begin{eqnarray}
H(G)&=&\sum_i [\theta_i k_i^{out}+\theta_i k_i^{in}+(\theta_0-\theta_i)k_i^\lr]\nonumber\\
&=&\sum_i \theta_i k_i+\theta_0 L^\lr
\end{eqnarray}
where we have defined $\theta_i\equiv
\theta_i^{in}=\theta_i^{out}=-\ln x_i$ and
$\theta_0\equiv-\ln\alpha$. The above expression highlights that the
constraints required in order to reproduce all the topological
properties of the WTW discussed so far are the undirected degree
sequence $\{k_i\}$ and the number of reciprocated links $L^\lr$, or
equivalently the reciprocity $r$. If the maximum likelihood
principle \cite{mylikelihood} is applied to this model, it is
straightforward to show that the parameters reproducing a given
snapshot of the WTW must be set to the particular values
$\alpha^*=e^{-\theta_0^*}$ and $\{x^*_i\}=\{e^{-\theta^*_i}\}$
satisfying the following $N+1$ coupled equations
\begin{eqnarray}
\langle k_i\rangle&=&\sum_{j\ne i}q_{ij}=
\sum_{j\ne i}\frac{[2+(\alpha^*)^2]x^*_i x^*_j}
{1+[2+(\alpha^*)^2] x^*_i x^*_j }=
k_i\qquad\forall i\\
\langle L^\lr\rangle&=&\sum_{i\ne j}r_{ji}p_{ij}=\sum_{j\ne i}\frac{(\alpha^*)^2x^*_i x^*_j}
{1+[2+(\alpha^*)^2] x^*_i x^*_j }=L^\lr
\end{eqnarray}
The first of the two expressions above indeed coincides with the
condition fixing the values of $\{w_i\}$ in the undirected
configuration model as in Equation (\ref{eq_likelihoodw}), under the
identification given by Equation (\ref{eq_reabsorb}). The second
expression allows to enforce any value of the reciprocity $r$ as
additional constraint, thanks to the extra parameter
$\alpha=e^{-\theta_0}$. Note that this graph ensemble is
stochastically symmetric under link reversal, since $H(A)=H(A^T)$.
Furthermore, since the real WTW is well reproduced by the
aforementioned graph ensemble, it is also stochastically symmetric
under link reversal.

We have therefore shown that the topology of the WTW for any year
$t$ since 1950 is completely reproduced by specifying the undirected
degree sequence $\{k_i(t)\}$ and the reciprocity $r(t)$. This
implies that, in order to explain why the WTW displays the structure
we observe, it is enough to explain these two topological
properties. However, while assessing the relevance of some
structural features is a rigorous procedure as we have shown so far,
explaining them in terms of underlying mechanisms involves a higher
degree of uncertainty and subjective interpretation. Bearing this in
mind, in what follows we suggest possible explanations for the two
structurally informative ingredients of the WTW topology. These
should be intended as candidate hypotheses rather than certain
mechanisms. Nonetheless, since the symmetry of the effect must be at
least that of the cause, the symmetry analysis carried out in the
preceding sections can be exploited to safely rule out explanations
that do not fulfill this principle.

\subsection{Topological Space and Embedding Spaces\label{sec_economic}}
We start by considering here the undirected degree sequence, while
in the next section we focus on the reciprocity. The accordance
between the configuration model and the undirected projection of the
WTW, {\em i.e.} the stochastic symmetry of the WTW under
transformations preserving the degree sequence, can be rephrased as
the finding that the network is the result of a random matching
process between the edges attached to every vertex. Vertices connect
to each other as the mere result of the constraint on their degrees.
The larger their degrees, the higher the probability that two
vertices are connected, with no higher-order effect on the topology.
This implies that, whatever the factor responsible for the observed
degree of a country, it must similarly respect the symmetry and be
such that \emph{the more a country is endowed with this factor, the
larger its degree and the higher its probability to connect to other
vertices, with no higher-order effect on the topology other than
those implied by the degree sequence}. If we denote the hidden
factor as $h$, and its value for vertex $i$ as $h_i$, the above
statement can be rephrased as \emph{the larger $h_i$, the larger the
expected value of $\langle k_i\rangle$}; similarly, \emph{the larger
$h_i$ and $h_j$, the larger the undirected connection probability
$q_{ij}$}. Therefore the hidden values $\{h_i\}$ must play exactly
the same role as that of the Lagrange multipliers $\{w_i\}$
controlling the expected degrees of all vertices in the undirected
configuration model as in Equation (\ref{eq_likelihoodw}). More in
general, the Lagrange multiplier $w_i$ could be a monotonic function
$f(h_i)$ of the hidden factor $h_i$. The above consideration
suggests at least two ways to test whether any empirically
observable quantity is indeed a good candidate as the hidden factor
determining the degree sequence in a given snapshot of the WTW.
First, one can solve the $N$ coupled equations in Equation
(\ref{eq_likelihoodw}) and obtain the set of values $\{w_i^*(t)\}$
which \emph{are} the exact values of the Lagrange multipliers
enforcing the observed degree sequence in year $t$, and then check
whether a candidate quantity $h(t)$, with empirically observed
values $\{h_i(t)\}$, is indeed in some approximate functional
dependence with these \linebreak multipliers, {\em i.e.} \be w^*_i(t)\approx
f[h_i(t),h_0(t)]\quad \forall i \label{eq_candidatew} \ee where $f$
can in general depend on, besides $h_i(t)$, a global time-dependent
parameter $h_0(t)$ setting the scale of the dependence. As a second
alternative, one could \emph{assume} the functional dependence
\linebreak $w_i(t)=f[h_i(t),h_0(t)]$, rewrite $q_{ij}(t)$ as \be
q_{ij}(t)=\frac{w_i(t)w_j(t)}{1+w_i(t)w_j(t)}=
\frac{f[h_i(t),h_0(t)]f[h_j(t),h_0(t)]}{1+f[h_i(t),h_0(t)]f[h_j(t),h_0(t)]}
\label{eq_candidateq} \ee and apply the maximum likelihood principle
to the resulting model, which now has only $h_0(t)$ as a free
parameter since the values  $\{h_i(t)\}$ are empirically accessible.
This leads to the single equation \be \langle
L^u(t)\rangle=\sum_{i<j}\frac{f[h_i(t),h^*_0(t)]f[h_j(t),h^*_0(t)]}{1+f[h_i(t),h^*_0(t)]f[h_j(t),h^*_0(t)]}=L^u(t)\quad\forall
i \label{eq_candidateL} \ee fixing the value of $h_0(t)$ for each
year $t$ and replacing Equation (\ref{eq_likelihoodw}). The goodness
of the assumed dependence can be tested by checking whether Equation
(\ref{eq_candidateq}), with the value $h^*_0(t)$ inserted in it,
reproduces the properties of the real network, in the same way as
Equation (\ref{eq_likelihoodw}) is used to assess the goodness of
the configuration model. Clearly, the first procedure is preferable
as it leaves the determination of the form of $f[h_i(t),h_0(t)]$ at
the end: once the values $\{w_i^*(t)\}$ are found exactly, one can
study the dependence of the latter on various candidate quantities
$h$, and with different functional forms. The second procedure
requires from the beginning the assumption one wants to test, and is
therefore less accurate; nonetheless, it could represent a further
test of the hypothesis if the output of the first method is used as
the input in the second one.

Both the approaches described above have been used to look for
hidden factors explaining the degree sequence, and consequently the
entire topology, of the undirected WTW \cite{mylikelihood,mywtw}.
The result of this analysis is that the Gross Domestic Product (GDP
in what follows) is a very good candidate factor. If $h_i(t)$ is
identified with the empirical GDP value of country $i$ in year $t$,
then an approximate linear relationship between $h_i(t)$ and the
value $w_i^*(t)$ obtained from Equation (\ref{eq_likelihoodw}) for
the same year is \linebreak observed \cite{mylikelihood}. This means that
Equation (\ref{eq_candidatew}) reduces to the simplest possible
functional form \be w^*_i(t)\approx h_i(t)\sqrt{h_0(t)}\quad \forall
i \ee where the proportionality factor has been denoted as
$\sqrt{h_0(t)}$ for convenience. This indicates that the probability
that a trade relationship (whatever its direction) exists between
countries $i$ and $j$ in year $t$ is \be q_{ij}(t)\approx
\frac{h_0(t)h_i(t)h_j(t)}{1+h_0(t)h_i(t)h_j(t)} \ee This result is
confirmed by assuming the above form of the connection probability,
using \linebreak Equation (\ref{eq_candidateL}) to find the value $h^*_0(t)$
generating the observed number of links, and checking that indeed
the empirical properties of the undirected WTW are reproduced
\cite{mywtw}. This result highlights that the larger (in economic
terms) a country, the higher its probability to connect to other
countries. According to our discussion at the end of Section
\ref{sec_undirected}, since the GDP is responsible for the degree
sequence of the WTW, it represents the symmetry-breaking variable
restricting the invariance of the network to equal-degree (or
similarly equal-GDP) equivalence classes. Contrary to what one could
expect on the basis of the spatial embedding of the WTW, no
significant dependence is found on other factors such as distance,
membership to common geographic areas or trade associations, {\em
etc}.

\begin{figure}[h]
\begin{center}
\includegraphics[width=.66\textwidth]{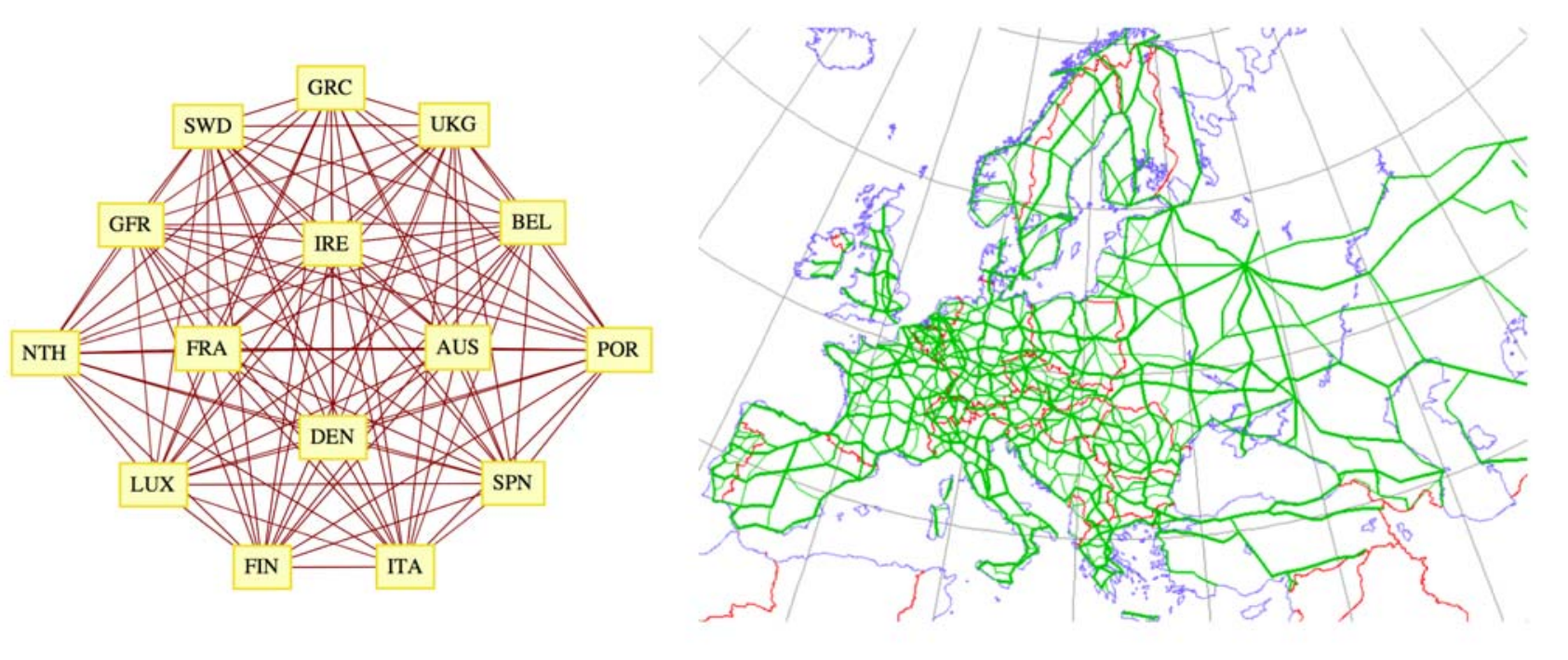}
\includegraphics[width=.3\textwidth]{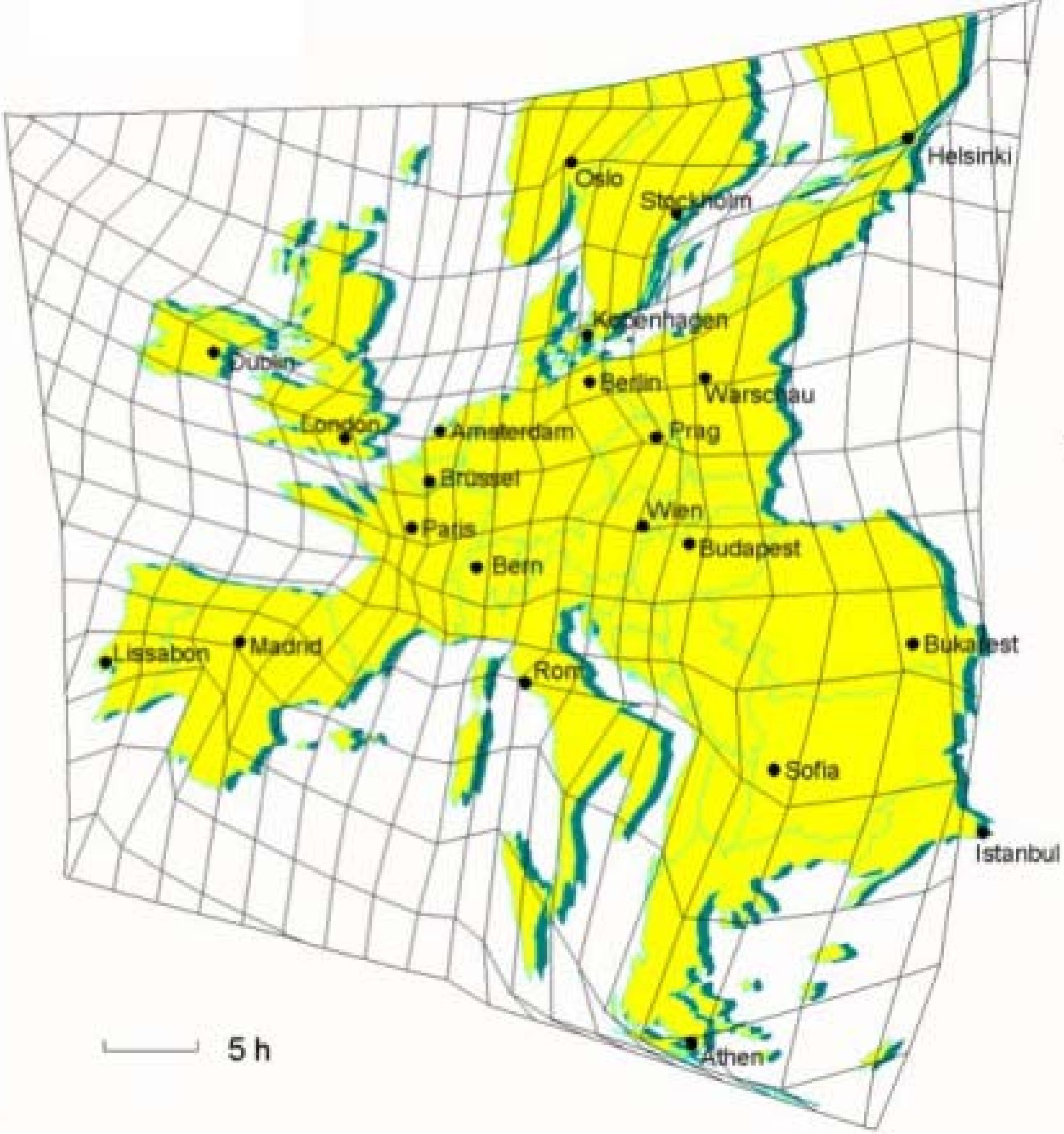}
\end{center}
\caption{European (EU-15) trade network as it would appear in  topological space (left panel), in Euclidean space assuming that trades travel mainly on the road network (middle panel, after \cite{road}), and after also taking into account the space modification due to traveling times (right panel, after \cite{times}).  \label{fig_map}}
\vspace{-0.4cm}
\end{figure}

The above result is very instructive in the light of the relation
between network structure and symmetry. What we should bear in mind,
when we consider symmetry breaking in the field of network theory,
is that symmetry (invariance) is hard to depict unless we use
analytical tools. Our imagination, intended as the faculty of
forming images, has been educated to depict shapes in  Euclidean
spaces. Whenever we must traduce shapes from Euclidean to
topological spaces, we are inevitably biased by the fact the we tend
to recall the Euclidean representation of forms in the new space.
This overlapping of spaces generates misrepresentation. To better
stress out the conundrum of spaces' inequality representation, in
Figure \ref{fig_map} we picture the trade network of Europe (EU-15),
as it would appear in  topological space (left panel) and in
Euclidean space (middle panel), assuming that trades travel mainly
on the road network \cite{road}. While the Euclidean representation
of the road network, except for the scale and a certain degree of
abstraction, is conformal to the real system's shape, the
corresponding representation of the trade network in a metric space
(the plane), is purely conventional. Indeed, we could have
represented the same network in several different ways, e.g.,
arraying in a circle or randomly scattering nodes.  We could
actually produce \emph{ad libitum} different Euclidean embeddings of
the same graph. The topological-Euclidean dichotomy is further
complicated by the fact the system represented by the WTW network is
immersed in an economic space, involving variables and relations in
large number, that are not detected in the network. Consider for
example the traveling time of goods, which is determined by several
exogenous factors. Traveling time, together with energy efficiency
and labor costs, is among the major factors affecting shipping
costs. In the right panel of Figure \ref{fig_map} we show a `metric
representation' of the space modification due to traveling time
\cite{times}. It is noteworthy that in an economic space distances
are not merely Euclidean and the compound metric is made by length,
time, labor costs and energy units at the minimum.

Indeed, the result that the WTW is excellently reproduced by a
connection probability that uniquely depends on the GDP indicates
that the space modification is even more extreme, as at a global
level geographic distances appear to play almost no role. In regular
lattices the overall permutation symmetry of vertices is broken by
positions in Euclidean space and is restricted to invariances of
lesser order such as translational symmetry \cite{symmetry1}. In
geographically embedded networks such as that shown in the middle
panel of Figure \ref{fig_map}, the irregularity of the geography
further restricts the symmetry properties. When additional variables
are also taken into account, even stronger distortions take place as
in the right panel of Figure \ref{fig_map}, and in the case of the
WTW we are in an extreme situation where the symmetry-breaking
variable is virtually only the GDP, and the distance dependence
practically disappears. The properties of the network must be
therefore interpreted in economic space rather than geographic
space. Still, in this space we find a remarkable symmetry: countries
with the same GDP are statistically \linebreak equivalent \cite{symmetry1}, and
pairs of countries with the same couple of GDP values have the same
probability to trade. In other words, we can rephrase the symmetry
properties of the WTW we discussed in \linebreak Section \ref{sec_undirected}
in terms of the GDP values rather than the degree sequence. This
invariance is preserved despite the heterogeneity of the GDP across
world countries increases in time \cite{myinterplay}, which means
that the intensity of the GDP-induced symmetry breaking also
increases. And, despite the latter determines ever-increasing
divergences between the values of the connection probability
$q_{ij}$ across pairs of countries, the average probability
$2\sum_{i< j}q_{ij}/N(N-1)$ remains almost constant as indicated by
the stationarity of the undirected connectance shown in Figure
\ref{fig_density}.

\subsection{The Reciprocation Process of World Trade and Spatial Symmetry Breaking\label{sec_revolution}}
As we mentioned, the second ingredient required in order to explain
the topological properties of the WTW is the reciprocity $r$, which
coincides with the conditional connection probability $r_0$ as
indicated by Equation (\ref{eq_rr}). While we have shown that the
marginal connection probability varies greatly among different pairs
of vertices, a property that can be traced back to the heterogeneous
degrees and possibly explained by the GDP values, the conditional
connection probability is uniform and must therefore be related to a
completely different mechanism. The heterogeneity of vertex degrees,
or of GDP values, is completely reflected in the marginal connection
probability while it is not reflected at all in the conditional
connection probability and in the reciprocity. To better understand
the problem, we now consider the temporal evolution of the
reciprocity and show how this may suggest possible explanations.

\begin{figure}[h]
\begin{center}
\includegraphics[width=.65\textwidth]{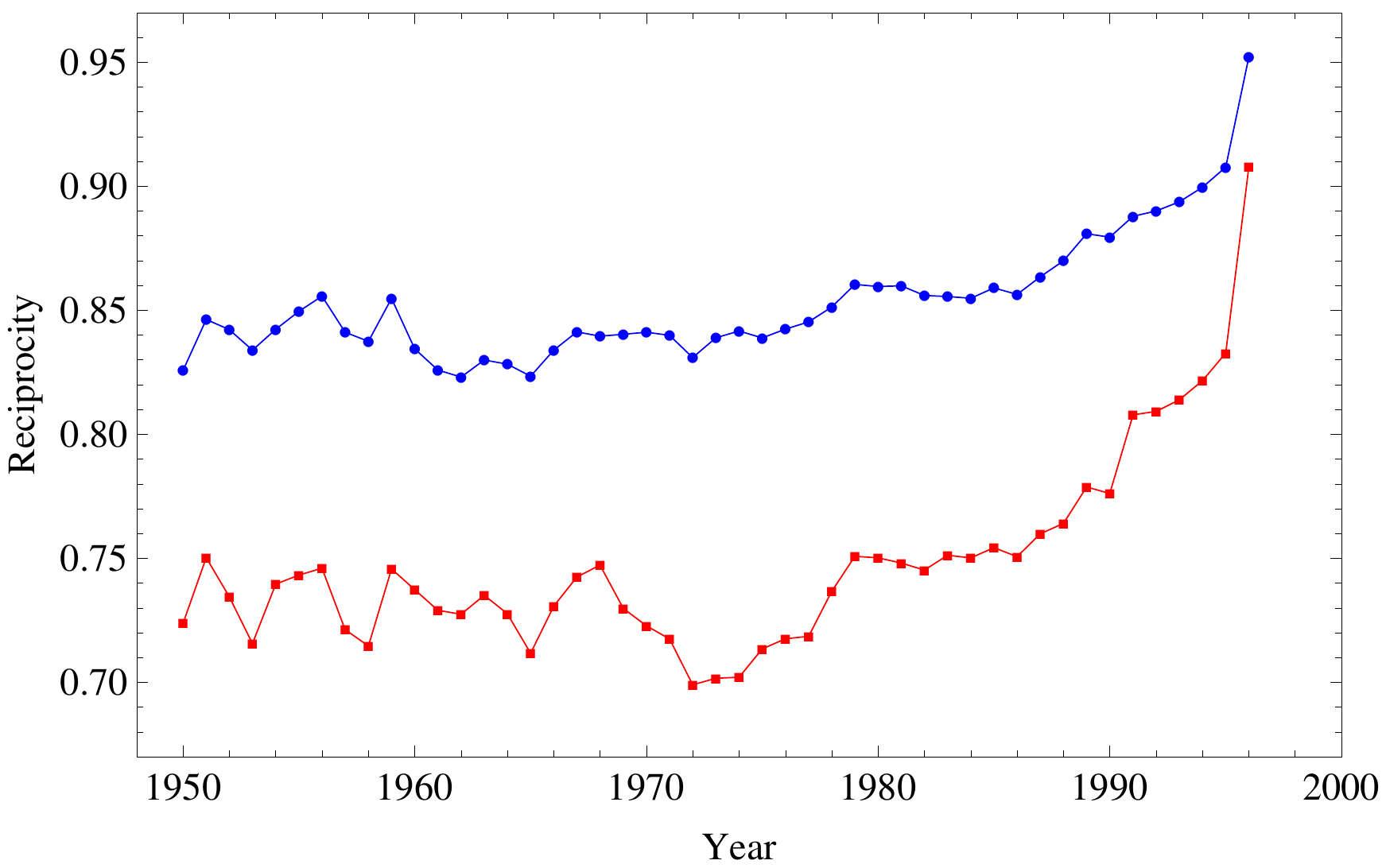}
\vspace{-0.5cm}
\end{center}
\caption{Evolution of the reciprocity measures $r(t)$ (blue points) and $\rho(t)$ (red points) in the directed version of the World Trade Web.\label{fig_reciprocity}}
\end{figure}

As clear from Equations (\ref{eq_kbothktot}) and (\ref{eq_kr0}), the
proportionality constant $r_0(t)/2$ between $k^\lr_i(t)$ and
$k^{tot}_i(t)$ is time-dependent. As Equation (\ref{eq_rr})
indicates, this means that the reciprocity $r(t)=r_0(t)$ of the
network must also change in time. In Figure \ref{fig_reciprocity} we
show the empirical evolution of $r(t)$. Indeed, we find that the
reciprocity of the WTW has evolved dramatically during the period
considered. In particular, we see that $r(t)$ has been fluctuating
about a constant value from 1950 to the late 1970's. Then, from the
late 1970's to the late 1990's, a steady increase of $r(t)$ took
place. More importantly, this occurred despite the density of
undirected trade relationships (the undirected connectance $\bar{b}$
shown in Figure \ref{fig_density}) remained approximately constant
during the same period. This indicates that, from the late 1970's
on, there has been an establishment of many new directed trade
relationships mainly between countries that had already been trading
in the opposite direction, rather than between countries that had
not been trading at all. That is, the reciprocation process of
unidirectional trade channels has dominated the formation of new
trade relationships between non-interacting countries.

The above results have implications for the evolution of the
symmetry properties of the network. As it was highlighted in the
previous sections, the equiprobability symmetry of the configuration
model allows a statistical interpretation of the real undirected
WTW.  Higher order topological variables can be explained by the
degree sequence. The invariance of the Hamiltonian is preserved
through time, as reflected by the stationarity of some topological
variables (Figure \ref{fig_density}). The stationarity is however
disrupted as we change lens and move from the undirected to the
directed graph. Reciprocity determines a clear symmetry breaking in
the directed analogue of the above invariance, as the in- and
out-degrees alone are no longer enough to characterise the network.
The intensity of this symmetry breaking evolves in time, as evident
in the trend of the reciprocity $r(t)$, and even more of $\rho(t)$.
Also, while the second type of link reversal symmetry (transpose
equiprobability) is approximately unchanged over time, since the
approximate equality $k^{in}_i(t)\approx k^{out}_i(t)$ holds
throughout the interval considered, the adjacency matrix suddenly
starts becoming \emph{more symmetric}: $A$ and $A^T$ become more and
more similar in time, indicating that the WTW has undergone a strong
evolution towards higher levels of link reversal symmetry of the
first type (transpose equivalence). The reason of such a sudden
change is obscure. The evolution may have been either driven by
other topological variables, {\em i.e.} it was endogenous to the
network, or determined by some hidden variables, thus exogenous to
the network. In other words:  the change either pertains thoroughly
the topological space or comes from an `outer embedding space',
where the exogenous variables belong to, that shapes the topological
space \cite{symmetry1}. Besides, it may also be possible that the
symmetry breaking actually occurred in  this latter space and
consequently affected the topological space. In what follows we
explore this problem in more detail.

A first natural explanation of the above empirical result could be
looked for in an overall increase in the number of directed trade
relationships during the period considered, a possibility consistent
with the globalisation process. Note that in our case Equation
(\ref{eq_cu}) implies that, while the undirected connectance
$\bar{b}(t)$ is approximately constant, the directed connectance
$\bar{a}(t)$ (hence the density of directed trade relationships) of
the WTW has indeed increased significantly due to the observed
increase of $r(t)$. In order to understand whether the observed
increase in reciprocity is merely due to the overall increase in
link density, it is important to recall our discussion in Section
\ref{sec_rho}, where we stressed the importance of using $\rho$
instead of $r$ since the former washes away density effects. Since
in this case the link density $\bar{a}(t)$ changes in time, using
$\rho(t)$ instead of $r(t)$ is also important in order to correctly
quantify the temporal evolution of the reciprocity. In
Figure \ref{fig_reciprocity}, besides $r(t)$, we also show the
evolution of $\rho(t)$. Unlike $r(t)$, the behaviour of $\rho(t)$ is
informative and clearly shows that the increase in density cannot
explain the increase in reciprocity. Remarkably, the evolution of
$\rho(t)$ is even more pronounced  than that of $r(t)$, indicating
that the change in density determines an underestimation of the
steep increase in reciprocity, if the latter is measured by $r$
rather than by $\rho$. The same consideration applies even if one
takes into account the fact that, according to our results discussed
above, the increase in the density of directed trade relationships
has occurred differentially across world countries, {\em i.e.} not
uniformly as in a directed random graph model with increasing
connection probability $p(t)$ but rather as in a directed
configuration model with heterogeneous probabilities $p_{ij}(t)$. If
the observed increase in reciprocity were merely due to a
differential, rather than homogeneous, increase of link density,
then we would observe $r_{ij}\approx p_{ij}$ as discussed in Section
\ref{sec_directed}. By contrast, the uniformity of $r$ rules out
this possibility. In other words, the inadequacy of the random graph
model and the configuration model in reproducing the observed
properties of the WTW rules out the possibility that the increase in
reciprocity is due to the globalisation process, at least the
component of the latter that is responsible for an (either
homogeneous or differential) increase in the density of directed
trade relationships.

As a second hypothesis, one could consider the establishment of new
trade agreements (preferentially between countries that had only
unidirectional trade relationships, and determining the
reciprocation of the latter) as a possible explanation for the
increase in the density of reciprocated links. However, trade
agreements do not explain the uniformity of the conditional
connection probability $r_{ij}(t)=r_0(t)$. For all years, the latter
is empirically found to be the same across all pairs of vertices,
which is in contrast with what expected from the formation of trade
agreements: an increased value of $r_{ij}(t)$ for pairs of countries
signing the agreement, determining an increased  heterogeneity of
$r_{ij}(t)$ across all pairs. Therefore the evolution in $r$ cannot
be explained by the formation of trade agreements. The uniformity of
the conditional connection probability also indicates that other
factors such as size, distance, etc. appear to be not enough in
order to explain how the reciprocity of world trade has evolved.

The above considerations show that the reciprocation of preexisting
unidirectional relationships appears to have occurred massively,
however with no preference for nearby or richer countries, and in a
way which cannot be traced back to an overall increase in the number
of trade relationships and trade agreements. We stress again that
all these factors must have had an impact on international trade
patterns, especially on the intensity of trade relationships,
however at a purely topological level they appear to be dominated by
a different mechanism, which is uniform across all pairs of
countries. In simplified terms, the evolution of the reciprocity of
the WTW could be approximated by a process where, with time-varying
but country-independent probability, a unidirectional trade
relationship existing at time $t$ becomes reciprocated in the
following year. Among the possible underlying mechanisms that could
generate this process, we must look for one displaying a temporal
trend which is synchronous with the one followed by the reciprocity
of the WTW and shown in Figure \ref{fig_reciprocity}. To this end,
it is useful to recall that in the case of the WTW, vertices and
links are samples of vertices and links of a larger underlying
network. Indeed, countries themselves do not trade; rather, firms
and consumers trade. Hence there are at least two submerged, and
much larger, networks: one of goods---final products---and one of
production factors---raw materials and semi-products (together with
a third network related to the service market). The WTW may be
considered as an overlapping map of these two networks. While the
two hypotheses advanced above mainly concerned the network of final
products, one could look for an explanation relative to the
production network (a network composed by factories as vertices and
productive means as links). The hypothesis that symmetry breaking
occurred in the economic space of the industrial sector in a period
starting between the 1970's to the early 1980's, with a significant
worldwide impact on the productive structure, has been recently
advanced \cite{7,8}. This transition was due on one hand to
decreasing energy costs of transport means and on the other hand to
raising labor costs. Firms therefore were stimulated to provide
production factors outside the division and began dispersing the
productive chain outside the company boundaries, sometimes abroad.
This process, named by economists \emph{outsourcing}, transformed
the space of firms from a Euclidean space, where providers where
separated from the production plant by physical distances, to an
economic space where physical distances were secondary to other
variables (i.e., changed the metric of economic space) \cite{8}.
This process was further reinforced by specialization and
technological enhancement, and was one of the driving forces of
globalisation. Note that, when a firm extends its productive chain
outside the national boundaries, new links may appear in the trade
network. This process can determine an increase in reciprocity, if
the new  countries entering the production process already import
from the firm's country. This mechanism can therefore provide a
candidate explanation, from the production side of network flows,
for the observed increase in reciprocity, which is also temporally
consistent with the empirical trend. If this hypothesis, which must
be further investigated, is correct, we would have faced a symmetry
breaking in economic space affecting topological space, partially
determining the phase transition we observe, and at the correct
moment in time.

\section{Conclusions}
In this paper we took full advantage of our investigation of the
symmetry properties of real \linebreak networks \cite{symmetry1} to
perform a detailed study in a specific case. We exploited our
concept of stochastic graph symmetries to introduce a new definition
of link reversal symmetry, {\em i.e.} transpose equiprobability. We
showed that, when combined with other symmetry properties of
directed networks, stochastic link reversal symmetry allows an
improved understanding of the reciprocity of real networks. In
particular, we have studied various symmetry properties of the World
Trade Web across its evolution. Our analysis also sheds a light in a
specific case on the symmetry properties of networks and symmetry
breaking in network topology. We found that a space-symmetry
approach to network theory may provide new insights into the complex
structure of the underlying system. We also made a conjecture about
the interplay between different spaces embedding the system,
captured by the topological space, that may lie behind some dramatic
changes observed in the detected topological variables. We believe
that spatial symmetry breaking deserves more attention as it may
lead to new perspectives in understanding complexity evolution and
specifically, those kind of transformations characterized by a
sudden leap in the complexity of the structure. Network theory
represents a theoretical framework that enables holistic analyses
and is suited to detect ongoing dynamics between the system's
components and the surrounding environment. In other words, network
theory is a paradigm  that considers the system as a whole and
distributes its functioning in space and time. The analysis of the
resulting symmetries carries information about the system.

More then twenty years ago, Marshall McLuhan, in the field of
communication theory, advocated the need to overcome the constrains
of conventional theories about communication, according to him
relegated to an Euclidean and `visual' space, to achieve a new
theory based on an ubiquitous and synchronous space, in his words:
an `acoustic space'.  He advanced the point that space (and not just
time) is an agent of communication. According to him, printed texts
have educated us to a sequential type of communication, whereas the
electronic age developed spatial communication: actors communicate
in the same time with the environment and mutually between them
\cite{9}. Nevertheless, in his opinion, communication theory did not
follow changes in communication media. In his own words: `The basis
of all contemporary Western theories of communication---the
Shannon-Weaver model---is a characteristic example of
left-hemisphere lineal bias. It ignores the surrounding environment
as a kind of pipeline model of a hardware container for software
content. It stresses the idea of inside and outside and assumes that
communication is a literal matching rather than making' \cite{10}.

We believe network theory is a paradigm intrinsically `spatial' and
`global', that best suits the need for a holistic theory versus a
sequential theory, that it could further benefit from the
interaction with other disciplines and concepts. We considered here
the case of symmetry and symmetry breaking, and showed that a
formalisation of the relation between these phenomena and network
properties is intriguing and informative, but at present still
incomplete. One of the present limitations is due to the fact that
the studied network is often a mere map of a larger, underlying
network, embedded in Euclidean or non-Euclidean spaces. Symmetry
breaking may occur in a different space, that is only indirectly
represented in the topological space. This indirect consequence
complicates a clear understanding of the underlying process, but
stochastic symmetries appear to capture patterns that are
unaccessible to exact symmetries. Future research must explore this
scenario more thoroughly, and possibly shed light on the relation
between network dynamics, symmetry breaking, the causal chain and
its premises.

\section*{Acknowledgements}
D.G. acknowledges financial support from the European Commission 6th
FP (Contract CIT3-CT-2005-513396), Project: DIME - Dynamics of
Institutions and Markets in Europe. We also thank Tiziano Squartini
for his help.

\bibliographystyle{mdpi}
\makeatletter
\renewcommand\@biblabel[1]{#1. }
\makeatother

\end{document}